\documentclass[]{iopart}
\usepackage[top=1.4in, bottom=1.4in, left=1.2in, right=1.2in]{geometry}
\usepackage{iopams}
\expandafter\let\csname equation*\endcsname\relax
\expandafter\let\csname endequation*\endcsname\relax
\usepackage{amsmath}
\usepackage{graphicx}
\usepackage{float}
\usepackage[hidelinks]{hyperref}
\usepackage{xcolor}
\hypersetup{
    colorlinks,
    linkcolor={red!50!black},
    citecolor={blue!50!black},
    urlcolor={blue!80!black}
}

\newcommand{\myfig}[4]{
\begin{figure}[htb]
  \hfill
  \centering
  \begin{minipage}[t]{0.49\columnwidth}
    \includegraphics[width=1\columnwidth]{#1}
  \end{minipage}
  \hfill
  \begin{minipage}[t]{0.49\columnwidth}
     \includegraphics[width=1\columnwidth]{#2}
  \end{minipage}
  \caption{#3}
  \hfill
  \label{#4}
\end{figure}
}

\newcommand{\bma}{\left( \begin{matrix}}
\newcommand{\ema}{\end{matrix} \right)}
\newcommand{\beq}{\begin{equation}}
\newcommand{\eeq}{\end{equation}}

\begin{document}
\title[]{Photoionization of few electron systems with a hybrid Coupled Channels approach}
\author{Vinay Pramod Majety, Alejandro Zielinski and Armin Scrinzi\footnote{Corresponding author: armin.scrinzi@lmu.de}}
\address{Physics Department, Ludwig Maximilians Universit\"at, D-80333 Munich, Germany}

\begin{abstract}
We present the hybrid anti-symmetrized coupled channels method for the calculation of fully differential
photo-electron spectra of multi-electron atoms and small molecules interacting with strong laser fields. 
The method unites quantum chemical few-body electronic structure with strong-field dynamics by solving the time 
dependent Schr\"odinger equation in a fully anti-symmetrized basis composed of multi-electron states 
from quantum chemistry and a one-electron numerical basis. Photoelectron spectra are obtained via the time dependent surface 
flux (tSURFF) method. Performance and accuracy of the approach are demonstrated for spectra from the helium and berryllium 
atoms and the hydrogen molecule in linearly polarized laser fields at wavelength from 21 nm to 400 nm. At long wavelengths, 
helium and the hydrogen molecule at equilibrium inter-nuclear distance can be approximated as single channel 
systems whereas beryllium needs a multi-channel description.
\end{abstract}

\vspace{2pc}

\maketitle

\section{Introduction}

Understanding laser - atom/molecule interaction has become an important research pursuit with the introduction of many versatile light probes over the past decade.
Promising experimental techniques like re-collision imaging \cite{Meckel13062008} and attosecond streaking \cite{Kienberger2004} are being pursued that aim to study time resolved
electron dynamics. One of the factors that always creates a certain amount of vagueness in interpreting these strong field ionization experiments is the possible
presence of multi-electron effects. An accurate interpretation of these experiments needs solutions of multi-electron time dependent Schr\"odinger equation (TDSE).
As perturbation theory is not valid in strong field regime, one resorts to direct numerical solutions of the TDSE. 

While simple single electron models or low dimensional models have been partially successful in explaining laser matter interactions, there have been several cases reported where
a more elaborate description of electronic structure becomes important. Some of the examples include inter-channel coupling leading to an enhancement in high harmonic generation (HHG)
from Xenon \cite{PhysRevLett.111.233005}, modification of angle resolved ionization yield of $CO_2$ \cite{PhysRevA.80.063411} and photoionization cross-sections in $SF_6$ \cite{JoseJ},
enhancement in HHG due to participation of doubly excited states in Beryllium \cite{PhysRevA.88.053412}, influence of nuclear motion \cite{PhysRevLett.96.143001}, presence of conical
intersections \cite{PhysRevLett.113.113003} and so on. All these instances need a more involved description of the electronic structure. 

With one and two electron systems, a full dimensional numerical treatment is possible in linearly polarized laser fields. For systems with more than 6 degrees of freedom a
full dimensional calculation is infeasible. There have been several efforts in the past decade to overcome this barrier of dimensionality for few electron systems by
choosing only a part of the Hilbert space that is seemingly important for the dynamics. Some of the approaches that are being employed are time dependent configuration
interaction methods \cite{PhysRevA.82.023406}, different variants of multi-configuration methods \cite{PhysRevA.71.012712,PhysRevA.86.053424, PhysRevA.87.062511,
PhysRevA.88.023402}, time dependent R-matrix method \cite{PhysRevA.79.053411}, and coupled channel methods \cite{PhysRevA.80.063411}. 

One of the observables that is typically measured in strong laser - atom/molecule interaction experiments is photoelectron spectra. While the methods listed above
\cite{PhysRevA.82.023406,PhysRevA.71.012712,PhysRevA.86.053424, PhysRevA.87.062511, PhysRevA.88.023402,PhysRevA.79.053411,PhysRevA.80.063411} have tried to include
multi-electron effects in photoionization studies, calculation of photoelectron spectra from multi-electron systems, especially at long wavelengths has remained out
of computational reach. The particular difficultly arises from the fact that, in order to compute photoelectron spectra the asymptotic part of the wavefunction is
required. This needs large simulation box volumes and access to exact single continuum states to project the wavefunction onto, at the end of time propagation.
Having large simulation boxes and computing single continuum states of a multi-electron system are expensive tasks, making these kind of computations costly. 

In this respect, a recently developed method called the time dependent Surface Flux (tSURFF) method \cite{Tao2012, Scrinzi2012} has turned out to be an attractive
solution. In the tSURFF approach, the wavefunction outside a certain simulation box is absorbed, and the electron flux through a surface before absorption is used to
obtain photoelectron spectra. This way photoelectron spectra can be computed with minimal box sizes.

We aim to deal with the difficulties of the few body problem and computation of photoelectron spectra by combining quantum chemistry with tSURFF for single electron
systems, through a coupled channels approach. Our coupled channels ansatz is similar in spirit to the one presented in \cite{PhysRevA.80.063411}. However, unlike in 
\cite{PhysRevA.80.063411}, we deal with anti-symmetrization exactly. We discretize our multi-electron wavefunctions with the ground state of the system and anti-symmetrized
products of the system's single ionic states and a numerical one-electron basis that we call the active electron basis. This ansatz is suitable to study single
ionization problems. We compute the ionic and neutral states for our basis using the state of the art quantum chemistry code COLUMBUS \cite{Lischka2011} giving us 
the flexibility to treat the ionic states at various levels of quantum chemistry. While the fully flexible active electron basis describes the ionizing electron, the
ionic basis describes the core polarization and the exact anti-symmetrization ensures indistinguishablity of the electrons. The inclusion of the field-free neutral
helps us to get the right ionization potential and start with the correct initial state correlation without much effort. We call our method hybrid fully anti-symmterized
coupled channels method and use the acronym haCC to refer to it in this work. Using tSURFF with haCC, we compute photoelectron spectra with minimal box sizes.

We intend to communicate in this article the mathematical formulation of our method, and demonstrate its usefulness by computing photoelectron spectra of $He$, $H_2$ and $Be$
in linearly polarized 21-400 nm wavelength laser fields and compare them with fully numerical two electron results. We discuss the advantages and limitations of such an approach 
through suitable examples.

\section{Mathematical formulation}
In this section, we describe our mathematical setup to solve the N-electron TDSE in the presence of an external laser field. We solve the TDSE 
\begin{equation} \label{eq:tdse}
 i \frac{\partial }{\partial t} \Psi = \hat{H} \Psi
\end{equation}
with fixed nuclei approximation and with dipole approximation which implies neglecting the spatial dependence of the laser field. Atomic units are used unless specified otherwise. The
non-relativistic N-electron field-free Hamiltonian can be written as:
\begin{equation}
 \hat{H} = \sum_{i} \left[ -\frac{1}{2} \nabla^2_i - \sum_p \frac{Z_p}{|\vec{r_i}-\vec{a}_p|} \right] + \sum_{ij;i<j} \frac{1}{|\vec{r}_i-\vec{r}_j|}
\end{equation}
where $Z_p$ is the nuclear charge and $\vec{a}_p$ are the nuclear coordinates of the $p^{th}$ nuclei. The interaction with the external laser field in length gauge 
is given by:
\begin{equation}
 \hat{D}_L = - \sum_j \vec{E}(t)\cdot\vec{r}_j
\end{equation}
and in velocity gauge by
\begin{equation}
 \hat{D}_V = \sum_j i \vec{A}(t)\cdot\vec{\nabla}_j .
\end{equation}

We describe our multi-electron discretization in detail in \ref{subsec:disc}, present the time propagation equations in \ref{subsec:tp} and the matrix 
elements in \ref{subsec:me}. The inverse of the overlap matrix can be computed efficiently using low rank updates which will be presented in \ref{subsec:oi}. Treating anti-symmetrization
exactly and including neutrals in our basis, introduces a technical difficulty in the form of linear dependencies in our basis. This is handled by performing a generalized inverse 
of the overlap matrix which will be presented in \ref{subsec:sing}. We work in mixed gauge for the reasons detailed in \cite{mixedg} and briefed in \ref{subsec:gauge}. Finally,
we present tSURFF for our coupled channels setup in section \ref{subsec:spec}.

\subsection{Multi-electron discretization} \label{subsec:disc}
We discretize our N-electron wavefunction with the quantum chemistry ground state neutral wavefunction and anti-symmetrized products of an ionic basis and a numerical one-electron basis. 
\begin{equation} \label{eq:ansatz}
 | \Psi (t) \rangle  \approx \sum_\mathcal{I} | \mathcal{I} \rangle C_{\mathcal{I}}(t) +  | \mathcal{G} \rangle C_\mathcal{G}(t)
\end{equation}
where
\begin{equation}
 | \mathcal{I} \rangle  = \mathcal{A} [ | i \rangle | I \rangle ] .
\end{equation}
Here, $\mathcal{A}$ indicates anti-symmetrization, $ |i\rangle $ are functions from a numerical one-electron basis that we call the active electron basis, $ | I \rangle $ and
$ | \mathcal{G} \rangle $ are (N-1) and N particle functions respectively and $C_{\mathcal{I}}(t)$, $C_\mathcal{G}(t)$ are the time dependent coefficients. 

The active electron is represented using finite elements for the radial coordinate and real spherical harmonics for the angular coordinates.
\begin{equation}
 | i(\vec{r}) \rangle = | f_i(r) \rangle | Y_{l_im_i}(\Omega) \rangle
\end{equation}
One each finite element we use high order scaled Legendre polynomials as basis functions. The typical orders we use are 10-14. The details of the finite element approach used here
can be found in \cite{PhysRevA.81.053845,helium}. A brief description is given in \ref{sec:apdx_fem} for the convenience of the reader.

We choose $ | I \rangle $ to be the eigenstates of the single ionic Hamiltonian obtained from the Multi-Reference Configuration Interaction 
Singles Doubles (MR-CISD) level of quantum chemistry. $|\mathcal{G}\rangle $ is chosen as the ground state of the system, also obtained from the MR-CISD level of quantum chemistry.
These quantum chemistry wavefunctions are constructed with an atom centered primitive gaussian basis as the starting point. While $|\mathcal{G}\rangle $ is the lowest eigenvector
of the N particle Hamiltonian with quantum chemistry, it is not the ground state of the Hamiltonian in our basis. Treating one of the electrons with the active electron basis that is
superior to the gaussian basis improves the ground state. 

The wavefunctions $ | \mathcal{G} \rangle $ and $ | I \rangle $ can be represented in a general form as sums of determinants:
\begin{equation}
 | I \rangle   = \sum_{p_1,p_2,..,p_{n-1}} d_{p_1,p_2,..,p_{n-1}} | \mathcal{A}[\phi_{p_1} \phi_{p_2} ... \phi_{p_{n-1}}] \rangle
\end{equation}
\begin{equation}
 | \mathcal{G} \rangle   = \sum_{p_1,p_2,..,p_{n}} d_{p_1,p_2,..,p_{n}} | \mathcal{A}[\phi_{p_1} \phi_{p_2} ... \phi_{p_{n}}] \rangle
\end{equation}
where $\phi_k$ are the Hartree-Fock orbitals of the neutral system. The same set of Hartree-Fock orbitals are used to construct both ionic and neutral CI functions. This allows us to use
simple Slater-Condon rules to compute any matrix elements between them. 

The explicit inclusion of the neutral ground state is motivated by the fact that, while the ionization process itself may be well described by one or few ionic channels, the initial
ground state may be correlated. In order to avoid inclusion of many ionic states just to describe the initial state, we include the neutral ground state explicitly, thereby reducing
the number of basis functions needed. This idea can be easily extended to include any specific correlated state that is of importance to a particular process. This gives the basis a
capability to be engineered based on any existing knowledge of the ionization process. In the current work, we only include the neutral ground state in our basis.

\subsection{Time propagation equations} \label{subsec:tp}
Substituting our ansatz (\ref{eq:ansatz}) into the TDSE (\ref{eq:tdse}) yields a set of coupled ordinary differential equations for the time dependent coefficients:
\begin{align}
 i  \left[ \langle \mathcal{G} | \mathcal{G} \rangle \frac{d C_{\mathcal{G}}}{dt} + \langle \mathcal{G} | \mathcal{I} \rangle \frac{d C_{\mathcal{I}}}{dt} \right]= 
 \langle \mathcal{G} | \hat{H} | \mathcal{G} \rangle C_{\mathcal{G}} + \langle \mathcal{G} | \hat{H} | \mathcal{I} \rangle C_{\mathcal{I}} \\
  i  \left[ \langle \mathcal{I} | \mathcal{G} \rangle \frac{d C_{\mathcal{G}}}{dt} + \langle \mathcal{I} | \mathcal{I} \rangle \frac{d C_{\mathcal{I}}}{dt} \right]= 
 \langle \mathcal{I} | \hat{H} | \mathcal{G} \rangle C_{\mathcal{G}} + \langle \mathcal{I} | \hat{H} | \mathcal{I} \rangle C_{\mathcal{I}} 
\end{align}

We time propagate the coefficients using an explicit fourth order Runge-Kutta method with an adaptive step size controller. In order to absorb the wavefunction at the box boundaries we
use infinite range Exterior Complex Scaling (irECS) \cite{PhysRevA.81.053845}. We typically choose simulation boxes larger than the spatial extent of the Hartree-Fock orbitals and
start absorption after the Hartree-Fock orbitals vanish. This implies that it suffices to complex scale only one of the N coordinates. 

The cost of time propagation scales with the number of ionic states (say $n_I$) as $n_I^2$ and it is independent of the exact number of electrons. This makes basis sets of kind
(\ref{eq:ansatz}) attractive for treating many electron systems.

\subsection{Matrix Elements} \label{subsec:me}
In order to solve the TDSE, we need to evaluate various operators in our chosen basis. Firstly, we introduce several generalized reduced density matrices that we need, with the help
of creation ($a^{\dagger}_k$) and annihilation ($a_k$) operators on the single particle state $\phi_k$. A $p$-th order generalized 
reduced density matrix between the (N-1) particle ionic functions is given by:
\begin{equation} \label{eq:rd}
 \rho^{IJ}_{k_1,...,k_p,l_1,...,l_p} = \langle I | a^{\dagger}_{k_1} ... a^{\dagger}_{k_p} a_{l_1} ... a_{l_p} | J \rangle
\end{equation}
Similarly, we define generalized Dyson coefficients between the N-particle neutral wavefunctions and (N-1) particle ionic wavefunctions as
\begin{equation} \label{eq:grd}
 \eta^{\mathcal{G}J}_{k_1,...,k_p,l_1,...,l_{p-1}} = \langle \mathcal{G} | a^{\dagger}_{k_1} ... a^{\dagger}_{k_p} a_{l_1} ... a_{l_{p-1}} | J \rangle
\end{equation}

With the help of these objects, we present below the final form of the matrix elements. The overlap matrix has the form
\begin{equation} \label{eq:ov}
 \begin{aligned}
 \langle \mathcal{G} | \mathcal{G} \rangle &= 1 \\
 \langle \mathcal{G} | \mathcal{I} \rangle &= \eta^{\mathcal{G}I}_{k} \langle \phi_k | i \rangle \\
 \langle \mathcal{I} | \mathcal{J} \rangle &= \langle i | j \rangle \langle I | J \rangle - \langle i | \phi_l \rangle \rho^{IJ}_{kl} \langle \phi_k | j \rangle \\
                                           &= \langle i | j \rangle \delta_{IJ} - \langle i | \phi_l \rangle \rho^{IJ}_{kl} \langle \phi_k | j \rangle
 \end{aligned}
\end{equation}
where $\eta^{\mathcal{G}I}_{k}$ can be identified with the Dyson orbital coefficients with respect to the Hartree-Fock orbitals and $\rho^{IJ}_{kl}$ are the
one particle reduced density matrices. 

The single particle operators have the form:
\begin{equation}
 \hat{T} = \hat{t}(1) + \hat{t}(2) + ... + \hat{t}(N)
\end{equation}
where $\hat{t}(u)$ is the single particle operator corresponding to the coordinate $u$. The matrix elements with respect to the basis (\ref{eq:ansatz}) take the form:

\begin{equation} \label{eq:1p_op}
\begin{aligned}
 \langle \mathcal{G} | \hat{T} | \mathcal{G} \rangle &= \rho^{\mathcal{G}\mathcal{G}}_{kl} \langle \phi_k | \hat{t} | \phi_l \rangle \\
 \langle \mathcal{G} | \hat{T} | \mathcal{I} \rangle &= \eta^{\mathcal{G}I}_{k} \langle \phi_k | \hat{t} | i \rangle + 
                                                                                    \eta^{\mathcal{G}I}_{klm} \langle \phi_k | \hat{t} | \phi_m \rangle \langle \phi_l | i \rangle \\
 \langle \mathcal{I} | \hat{T} | \mathcal{J} \rangle &= \langle i | j \rangle \langle I | \hat{t} |J \rangle + \langle i | \hat{t} | j \rangle \langle I |J \rangle
                                                         - \langle i | \hat{t} | \phi_l \rangle \rho^{IJ}_{kl} \langle \phi_k | j \rangle  \\
                                                         & \qquad - \langle i | \phi_l \rangle \rho^{IJ}_{kl} \langle \phi_k | \hat{t} | j \rangle 
                                                         - \langle i | \phi_c \rangle \langle \phi_a | j \rangle \langle \phi_b | \hat{t} | \phi_d \rangle \rho^{IJ}_{abcd} 
\end{aligned}
\end{equation}
where $\eta^{\mathcal{G}I}_{klm}$ are the three index generalized Dyson coefficients, Eq.(\ref{eq:grd}), and $\rho^{IJ}_{abcd}$ are the two particle reduced density matrices, Eq. (\ref{eq:rd}). 

Finally, the two particle operators have the form:
\begin{equation}
 \hat{V}^{(2)} = \sum_{ij;i<j} \hat{v}(ij)
\end{equation}
with the matrix elements

\begin{equation}
 \begin{aligned}
 \langle \mathcal{G} | \hat{V}^{(2)} | \mathcal{G} \rangle &= \frac{1}{2} \rho^{\mathcal{G}\mathcal{G}}_{abcd} \langle \phi_a \phi_b | \hat{v} | \phi_c \phi_d \rangle \\
 \langle \mathcal{G} | \hat{V}^{(2)} | \mathcal{I} \rangle &= \eta^{\mathcal{G}I}_{klm} \langle \phi_k \phi_l | \hat{v} | \phi_m i \rangle + \frac{1}{2} \eta^{\mathcal{G}I}_{abcde} \langle \phi_a \phi_b | \hat{v} | \phi_d \phi_e \rangle \langle \phi_c | i \rangle \\
 \langle \mathcal{I} | \hat{V}^{(2)} | \mathcal{J} \rangle &= \frac{1}{2} \rho^{\mathcal{I}\mathcal{J}}_{abcd} \langle \phi_a \phi_b | \hat{v} | \phi_c \phi_d \rangle \langle i | j \rangle
                                                              + \underbrace{\rho^{IJ}_{kl} \langle \phi_k i | \hat{v} | \phi_l j \rangle}_{\text{Direct term}}
                                                              - \underbrace{\rho^{IJ}_{kl} \langle \phi_k i | \hat{v} | j \phi_l \rangle}_{\text{Standard exchange term}} \\ 
                                                              & \;\;\; - \underbrace{\rho^{IJ}_{abcd} \langle \phi_a i | \hat{v} | \phi_c \phi_d \rangle \langle \phi_b | j \rangle 
                                                              -\rho^{IJ}_{abcd} \langle \phi_a \phi_b | \hat{v} | \phi_c j \rangle \langle i | \phi_d \rangle
                                                              - \frac{1}{2} \rho^{IJ}_{abcdef} \langle \phi_a \phi_b | \hat{v} | \phi_d \phi_e \rangle \langle i | \phi_f \rangle \langle \phi_c | j \rangle
                                                              }_{\text{Other exchange terms due to non-orthogonality}}
\end{aligned}
\end{equation}

where $\eta^{\mathcal{G}I}_{abcde}$ are the five index generalized Dyson coefficients, Eq.(\ref{eq:grd}) and $\rho^{IJ}_{abcdef}$ are the three particle reduced density matrices,
Eq. (\ref{eq:rd}). 

In order to compute the two-electron integrals, we first project the Hartree-Fock orbitals onto a single center expansion:
\begin{equation}
 \phi_k(\vec{r}) = \sum_{q_kl_km_k} d_{q_kl_km_k} Y_{l_km_k}(\theta,\phi)
\end{equation}
where $q_k$ are some radial quadrature points, $l_k,m_k$ refer to the angular momentum functions and use these expansions with the multi-pole expansion:
\begin{equation}
 \frac{1}{|\vec{r}_1-\vec{r}_2|} = \sum_{LM} \frac{4\pi}{2L+1} \frac{r^L_{<}}{r^{L+1}_{>}} Y_{LM}(\theta_1,\phi_1) Y^{*}_{LM}(\theta_2,\phi_2)
\end{equation}
with $r_< = \text{min}(r_1,r_2)$ and $r_> = \text{max}(r_1,r_2)$. These two particle operators pose a challenge for efficient computation. While the direct term is relatively easy to
handle, the exchange terms consume a major portion of the Hamiltonian setup time. 

\subsection{Inverse of the overlap matrix} \label{subsec:oi}
The overlap matrix (\ref{eq:ov}) does not have the form of an overlap of a standard finite element basis. The overlap of a standard finite
element basis has a banded structure that usually helps in an efficient computation of the inverse. Non-orthogonality between the active electron
basis and the Hartree-Fock orbitals leads to extra cross terms that destroy the banded structure in general and complicate the computation of the 
inverse. However, the inverse of the overlap can still be computed efficiently using low rank updates. We use here the Woodbury formula \cite{woodbury} 
to compute our inverse, according to which, the inverse of a modified matrix of the form $(S_0 - U \Lambda U^\dagger)$ can be computed as:
\begin{align} \label{eq:wbf}
 S^{-1} &= (S_0 - U \Lambda U^\dagger)^{-1} \nonumber \\ &= S_0^{-1} - S_0^{-1} U (U^{\dagger} S_0^{-1} U - \Lambda^{-1})^{-1} U^{\dagger} S_0^{-1} 
\end{align}
As an example with 2 ionic states and 1 neutral, we show that the overlap matrix (\ref{eq:ov}) can be cast in the form:  
\begin{equation}
S=
\underbrace{\bma s_0 & 0 & 0 \\ 0 & s_0 & 0 \\ 0 & 0& 1\ema}_{S_0}
-\underbrace{\bma  u & 0 & 0 \\ 0 & u & 0 \\ 0 & 0& 1\ema}_{U}
\underbrace{\bma 
\rho^{11} & \rho^{12} & \eta^{\mathcal{G}1}\\
\rho^{21} & \rho^{22} & \eta^{\mathcal{G}2}\\
[\eta^{\mathcal{G}1}]^T & [\eta^{\mathcal{G}2}]^T&0
\ema}_{\Lambda}
U^\dagger
\end{equation}
which resembles the left hand side of the Woodbury formula (\ref{eq:wbf}). Here, $(s_0)_{ij}=\langle i| j \rangle$ and $u_{ik}=\langle i | \phi_k\rangle$.
If $n_i$ is the number of active electron basis functions and $n_{hf}$ be the number of Hartree-Fock orbitals that is much smaller that $n_i$, then the dimensions
of $s_0$ are $n_i\times n_i$ and the dimensions of matrix $u$ are $n_i\times n_{hf}$.

Let $n_I$ be the number of ionic states in general. Then the dimensions of the overlap matrix $S$ and $S_0$ are $(n_I n_i + 1) \times (n_I n_i + 1) $.
The dimensions of the matrix $U$ are $ (n_I n_i + 1) \times (n_I n_{hf} + 1) $ and the matrix $\Lambda$ are $ (n_I n_{hf} + 1) \times (n_I n_{hf} + 1) $. This
low rank structure of the extra correction terms can be utilized to compute the inverse efficiently by using the Woodbury formula.

\subsection{Handling linear dependencies} \label{subsec:sing}
 Anti-symmetrization and non-orthogonality of the active electron basis with respect to the Hartree-Fock orbitals may render our basis linearly dependent. 
 If the $\{|i\rangle\}$-basis is near-complete w.r.t.\ the HF-oribtal basis
 \begin{equation}
  \langle \phi_k | i \rangle \; [s_0^{-1}]_{ij} \; \langle j | \phi_l \rangle \approx \delta_{kl} ,
 \end{equation}
 it is possible to find coefficients $c_{i,I}$ such that
 \begin{equation}
  \sum_{i,I} c_{i,I}  \mathcal{A}[|i\rangle |I\rangle] \approx 0 .
 \end{equation}
 A simple case when this can happen is, if one and the same HF orbital appears in all the ionic determinants, 
which renders all neutral determinants involving the complete basis =0.
This makes the overlap matrix non-invertible. A possible solution would be to orthogonalize the active
 electron basis with respect to the Hartree-Fock orbitals. But this is not an easily implementable solution with a CI ionic basis. For each determinant, the set of Hartree-Fock orbitals with
 respect to which the active electron basis must be orthogonal is different. 

As an alternative solution we use a generalization of the Woodbury formula (\ref{eq:wbf}) to compute the
inverse of a matrix only on the subspace of the non-zero eigenvectors of the matrix.
Let $D$ denote the  $n_0\times n_z$ matrix of eigenvectors with near-zero eigenvalues $z_p<\epsilon, p=1,..., n_z$ of the generalized 
eigenvalue problem
\beq
SD = S_0D z,
\eeq
with $z$ denoting diagonal matrix of the eigenvalues $z_p$ and $D$ satisfying the orthonormality relation $D^{\dagger}S_0D = \mathbf{1}$. In general,
there will be comparatively few such eigenvectors $n_z \ll n_0$ that can be easily determined by an iterative solver. We can remove these singular
vectors from our calculation by the projector
\beq
Q = 1-DD^\dagger S_0.
\eeq
The projector property $Q^2=Q$ can be easily verified. As the projector refers to the generalized eigenvalue problem with $S_0\neq \mathbf{1}$,
$Q$ is not an orthogonal projector, that is $Q^\dagger\neq Q$. We define a pseudo-inverse $\tilde{S}^{-1}$ of the restricted $S$ on the 
subspace of generalized eigenvectors with non-zero eigenvalues with the property
\beq\label{eq:pseudoInverse}
\tilde{S}^{-1} S Q = Q.
\eeq
One can verify directly that the generalized Woodbury formula
\beq
\tilde{S}^{-1} = QS_0^{-1}\left[1-U(U^\dagger QS_0^{-1}U-\Lambda)^{-1}U^\dagger QS_0^{-1}\right]
\eeq
satisfies the definition (\ref{eq:pseudoInverse}). The matrix $(U^\dagger QS_0^{-1}U-\Lambda)$ is invertible
on all vectors appearing in $U^\dagger Q$ to its right, as exactly the singular vectors are removed by 
the projector $Q$. Apart from the necessity to determine $D$ during setup,
the correction does not significantly increase the operations count for the inverse overlap. 

\subsection{Choice of gauge} \label{subsec:gauge}
In \cite{mixedg}, we had shown that when an electron is treated with a restricted basis, for example, in terms of a few bound states, the length gauge is a more natural gauge.
Compared to pure velocity gauge, the coupled channel computations converge quickly in mixed gauge with length gauge spanning the region of the ionic states and velocity gauge
thereafter for asymptotics. In this current work, we use continuous gauge switching, detailed in \cite{mixedg}, for its easy implementation. We use the resulting TDSE after
the length gauge form is transformed using the following transformation.
\begin{equation}
    U_c= 
\begin{cases}
    1& \text{for } r\leq r_g\\
    \exp\left[ i \vec{A}(t) \cdot \sum\limits_{j=1}^{N} \hat{r_j} \left( r_j - r_g \right) \right] & \text{for } r > r_g
\end{cases}
\end{equation}
Here, $r_g$ is the gauge radius that separates the length gauge and velocity gauge regions. 

\subsection{Computation of photoelectron spectra} \label{subsec:spec}

Computation of photoelectron spectra is expensive for two reasons. (1) The asymptotic part of the wavefunction is needed to extract photoelectron spectra, which means large simulation boxes
to preserve the asymptotic part and to avoid any numerical reflections that may corrupt the wavefunction. (2) Single continuum states are needed into which the wavefunction must be decomposed,
in order to obtain photoelectron spectra. These two problems are circumvented in a recently developed method tSURFF\cite{Tao2012, Scrinzi2012} by computing photoelectron spectra, through a time integration
of electron flux flowing through a surface defined by a radius $R_c$ called the tSURFF radius. The Coulomb potential is smoothly turned off before $R_c$, which implies that the 
scattering solutions thereafter are well known Volkov solutions. $R_c$ becomes a convergence parameter, and by varying this radius, one can compute spectra to a given accuracy. 
This method has been explained in detail in previous works for single ionization in \cite{Tao2012} and for double ionization in \cite{Scrinzi2012}. A proposal for extension of this method for
single ionization of multi-electron systems has been outlined in \cite{Tao2012}. We describe here the application of the method with our coupled channels setup.

Let $\chi_k$ be the scattering solutions which take the form of Volkov soltutions beyond $R_c$ and $\Psi(T)$ be the wavefunction at some large time T. According to tSURFF for single electron
systems, photoelectron spectra can be computed as $\sigma_k = |b_k|^2$ with $b_k$ defined as:
\begin{align}
 b_k &= \langle \chi_k (T) |  \Theta(R_c) | \Psi(T) \rangle \nonumber \\ &= i \int_0^T dt \langle \chi_k(t) | \left[ -\frac{1}{2} \triangle + i \vec{A}(t)\cdot \vec{\nabla} , \Theta(R_c) \right] | \Psi(t) \rangle
\end{align}
Here $\Theta(R_c)$ is a Heaviside function that is unity for $r > R_c$ and 0 elsewhere.

This formulation can be easily extended to the N electron problem in a coupled channels setup. In this setup, we mostly take a set of ionic bound states for the ionic basis. These states
have a finite extent. We may choose $R_c$ such that the electrons described by the ionic basis vanish by $R_c$ which means all the exchange terms in the Hamiltonian vanish
after $R_c$. The remaining direct potential can be smoothly turned off just as the Coulomb potential. This implies that the wavefunction beyond $R_c$ evolves by the Hamiltonian:
\begin{equation}
 H (r>R_c) = H_{ion} \otimes \hat{1} + \hat{1}_{ion} \otimes \left[ -\frac{1}{2} \triangle + i \vec{A}(t)\cdot \vec{\nabla} \right]
\end{equation}
that allows for a complete set of solutions of the form:
\begin{equation} 
 \xi_{c,k}(\vec{r}_1,...,\vec{r}_n,t) = \mathcal{A} \left[ \kappa_c(\vec{r}_1,...,\vec{r}_{n-1},t) \otimes \chi_k(\vec{r}_n,t) \right]
\end{equation}
where $H_{ion}$ is the single ionic Hamiltonian and $\kappa_c(t)$ are time dependent ionic channel functions solving the TDSE
\begin{equation} \label{eq:ion_tdse}
 i\frac{\partial \kappa_c(t)}{\partial t} = \hat{H}_{ion} \kappa_c(t)
\end{equation}
within the ansatz in terms of field-free ionic states 
\begin{equation} \label{eq:ion_bas}
 |\kappa_c(t)\rangle = \sum_I | I \rangle d_{cI}(t).
\end{equation}
With the help of the $\xi_{c,k}$, channel resolved photoelectron spectra can be computed as
\begin{align}
 \sigma_{c,k} = | \langle \xi_{c,k}(\vec{r}_1,...,\vec{r}_n,T) |  \Theta(R_c) | \Psi(\vec{r}_1,...,\vec{r}_n,T) \rangle |^2 
\end{align}
and the asymptotic decomposition of $\Psi$ in terms of $\xi_{c,k}$ is obtained as 
\begin{equation} \label{eq:cc-tsurff}
\begin{aligned}
&\langle \xi_{c,k}(\vec{r}_1,...,\vec{r}_n,T) |  \Theta(R_c) | \Psi(\vec{r}_1,...,\vec{r}_n,T) \rangle \\
              & = i \int_0^T dt \langle \mathcal{A} \left[ \kappa_c(\vec{r}_1,...,\vec{r}_{n-1},t) \otimes \chi_k(\vec{r}_n,t) \right] | \left[ -\frac{1}{2} \triangle_n + i \vec{A}(t)\cdot \vec{\nabla}_n , \Theta_n(R_c) \right] | \Psi(\vec{r}_1,...,\vec{r}_n,t) \rangle \\
              & = i \int_0^T dt \langle \chi_k(\vec{r}_n,t) | \left[ -\frac{1}{2} \triangle_n + i \vec{A}(t)\cdot \vec{\nabla}_n , \Theta_n(R_c) \right] | \zeta_c(\vec{r}_n,t) \rangle
\end{aligned}
\end{equation}
where 
\begin{equation}
 \zeta_c(\vec{r}_n,t) = \langle \kappa_c(\vec{r}_1,...,\vec{r}_{n-1},t) |  \Psi(\vec{r}_1,...,\vec{r}_n,t) \rangle
\end{equation}
are Dyson-like orbitals. The commutator of the derivatives with the Heaviside function $\Theta$ gives $\delta$-like terms involving values and derivatives of $\Psi$ at
the surface $|\vec{r}|=R_c$. As we choose $R_c$ such that the Hartree-Fock orbitals vanish by then, we don't need to consider the exchange terms in computing $\zeta_c$. Along with
time propagating the N electron problem, one needs to also time propagate the ionic problem (\ref{eq:ion_tdse}).

\subsection{Spin symmetry}
As we solve the non-relativistic TDSE, the total spin of the system is conserved during the time evolution. We can therefore remove the spin degree of freedom through suitable linear
combinations of the anti-symmetrized products in the basis (\ref{eq:ansatz}) to enforce a particular spin symmetry. This reduces the size of our basis. We consider only singlet spin symmetric
systems in this work. As an example, we show how singlet spin symmetry can be enforced. Let $\uparrow$ and $\downarrow$ indicate the spin states $\pm \frac{1}{2}$ associated with a spatial
function. Choosing linear combinations of the kind:
\begin{equation}
 \mathcal{A} \left[ |I\rangle |i\rangle \right] :=  \frac{\mathcal{A} \left[ |I^{\uparrow}\rangle |i^{\downarrow}\rangle \right] 
                                                   - \mathcal{A} \left[ |I^{\downarrow}\rangle |i^{\uparrow}\rangle \right]}{\sqrt{2}}
\end{equation}
enforces singlet symmetry. This can be extended to creating linear combinations that enforce an arbitrary spin symmetry.

\section{Two-electron benchmark calculations}

We use two-electron full dimensional calculations (full-2e) as benchmark for our haCC computations. We solve the two-electron TDSE using an independent particle basis of the form:
\begin{equation} \label{eq:2e_exp}
 \Psi(\vec{r}_1,\vec{r}_2,t) = \sum_{k_1k_2l_1l_2m} c_{k_1k_2l_1l_2m}(t) f_{k_1}(r_1) f_{k_2}(r_2) Y_{l_1m}(\theta_1,\phi_1) Y_{l_2-m}(\theta_2,\phi_2)
\end{equation}
where $c_{k_1k_2l_1l_2m}(t)$ are the time dependent coefficients, $f_{k_1}(r_1),f_{k_2}(r_2)$ are functions from a finite element discretization of the same type as for our
active electron basis and $Y_{lm}$ are spherical harmonics. We use the same type of single center expansion for all the benchmark computations. A complete description of this
method will be presented elsewhere \cite{helium}. Solving the TDSE with the expansion (\ref{eq:2e_exp}) needs much larger computational resources compared to the haCC approach.

\section{Single photoelectron spectra}
In this section, we present photoelectron spectra from helium and beryllium atoms and from the hydrogen molecule with linearly polarized laser fields computed with the above described
coupled channels formalism. We also present the single photon ionization cross-sections for the beryllium atom and wavelength dependence of ionization yield for the hydrogen molecule
to compare with other existing calculations. We use $\cos^2$ envelope pulses for all the calculations and the exact pulse shape is given as
\begin{equation} \label{pulse}
 A_z(t) = A_0 \cos^2(\frac{\pi t}{2cT}) \; \sin(\frac{2\pi t}{T} + \beta)
\end{equation}
\begin{equation}
  E_z(t) = -\frac{dA_z(t)}{dt}
\end{equation}
where $A_0$ is the peak vector potential, $T$ is the single cycle duration, $c$ is the number of laser cycles and $\beta$ is the carrier envelope phase. We compare our results for helium and
the hydrogen molecule with full-2e calculations \cite{helium} and for beryllium with effective two electron model calculations. 

The convergence of the benchmark calculations and the haCC calculations were done systematically and independently. All the spectra presented here were computed with simulation box sizes on
the scale of 30-50 a.u. The radial finite element basis consisted of high order polynomials typically of orders 10-14 and the total number of radial basis functions was such that there
were 2-3 functions per atomic unit. The angular momenta requirement strongly depends on the wavelength. The longer wavelengths needed larger number of angular momenta for convergence.

\subsection{Helium} \label{sec:helium}
Helium is the largest atom that can be numerically treated in full dimensionality. With linearly polarized laser fields, the symmetry of the system can be used to reduce 
the problem to 5 dimensions. The energies of helium ionic states are $-2n^{-2}$ for principal quantum number $n$. The first two ionic states are separated by 1.5 a.u
in energy, which is large, for example, compared to a photon energy of 0.456 a.u at 100nm. This has been a motivation to treat helium as an effective single electron system
with XUV and longer wavelengths in some earlier works, for example in \cite{PhysRevA.89.021402}. We examine below, the validity of treating helium as an effective single electron system,
by comparing haCC calculations with full dimensional calculations at different wavelengths.

\myfig
{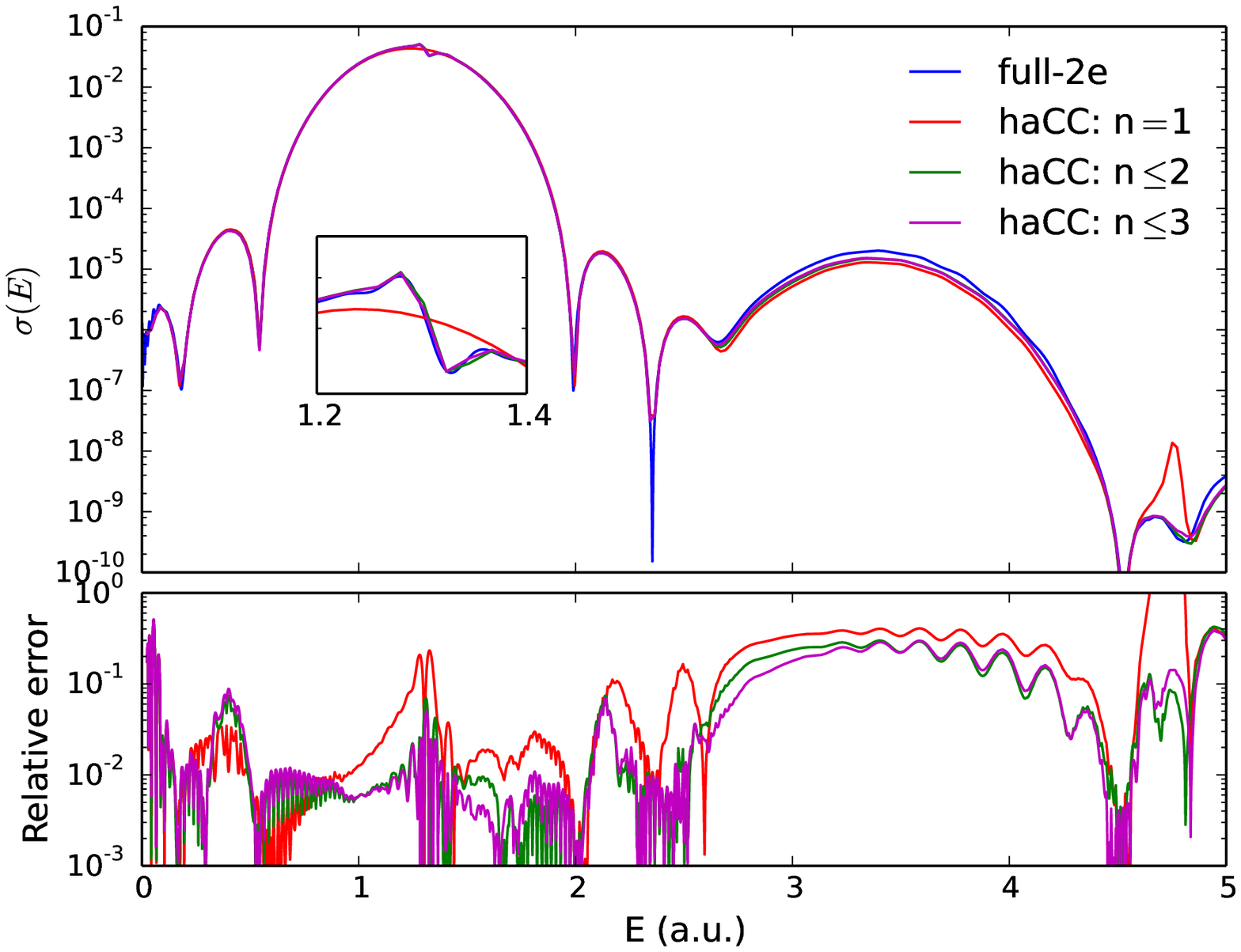} {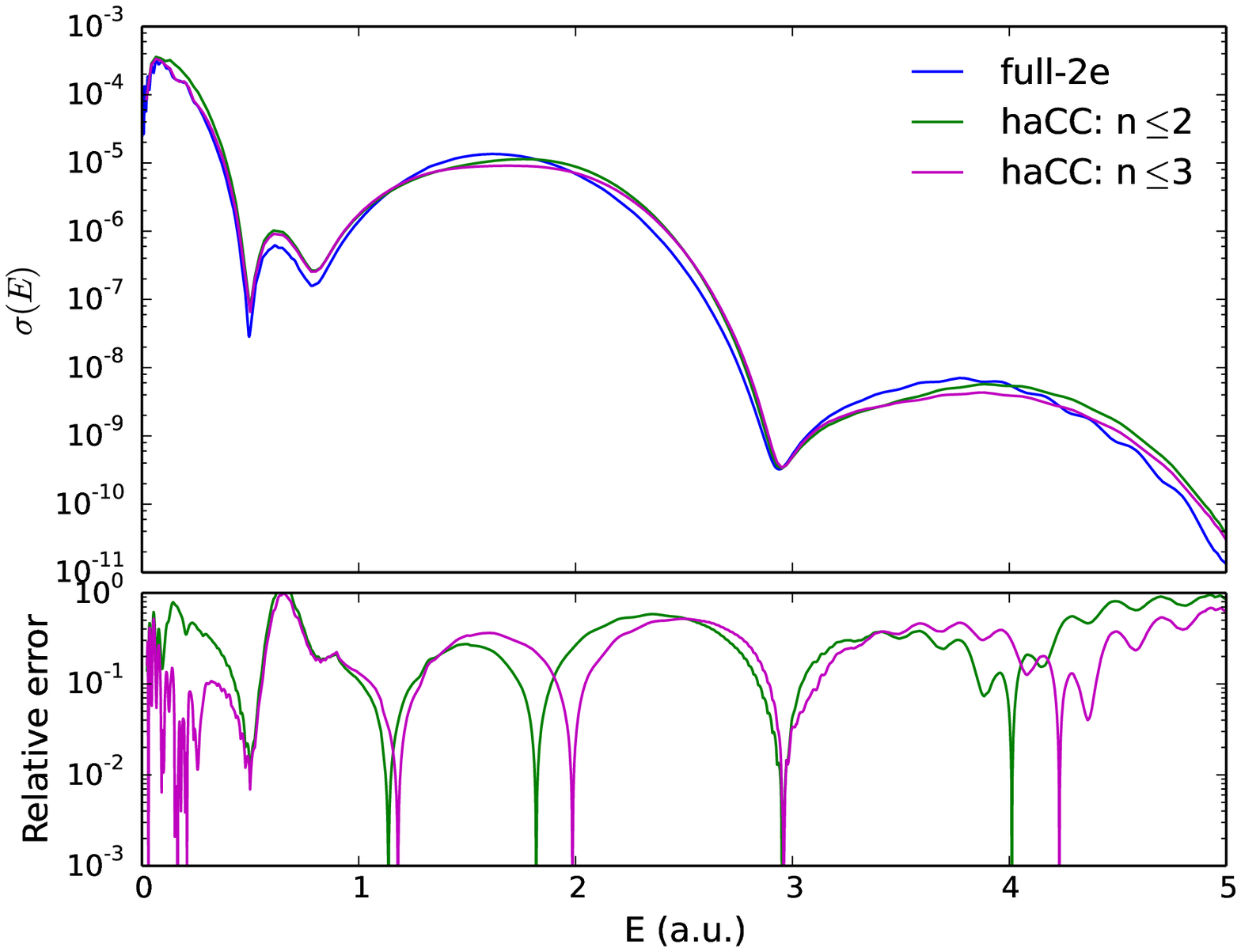}
{ Photoelectron spectra from helium with 3-cycle, 21nm laser pulse with a peak intensity of $10^{15} W/cm^2$. Left figure: ground state channel ($1s$) , Right figure: a first excited
state channel ($2p_z$). The upper panels show spectra obtained with a full-2e and haCC calculations with different number of ionic states included as indicated in the legend. Here, n is the 
principal quantum number. The lower panels show relative errors of haCC calculations with respect to full-2e calculations. The inset shows the $2s 2p$ resonance (see main text).}
{fig:He:21nm}

Figure \ref{fig:He:21nm} shows photoelectron spectra from helium with a 21 nm ($\omega$ = 2.174 a.u), 3 cycle laser pulse with a peak intensity of $10^{15} W/cm^2$. The one and two photon
ionization peaks of $1s$ and $2p_z$ channel spectra are shown. The relative errors of haCC calculations are computed with respect to the full dimensional calculation. The single photon
peak of the $1s$ channel is computed to a few percent accuracy, except for a feature around 1.3 a.u., with a single ionic state. The resonant feature can be identified with the $2s 2p$
doubly excited state \cite{Svatopluk}, which is reproduced to few percent accuracy with the addition of 2nd shell ionic states. While the position of the resonance is reproduced accurately
in the calculations presented here, the propagation time was well below the life-time of this resonance which is reflected in the width of the feature that is well above the natural
line width. It should be emphasized that the exact width emerges when propagation times are extended to large times comparable to the decay times of the doubly excited state. The two
photon peak of the $1s$ channel and the $2p_z$ channel spectra need more than a single ionic state and they could be computed only up to 15\% accuracy even after inclusion of 9 ionic
states ($n \leq 3$).

A broadband (few cycle) XUV pulse tends to excite the initial state into a band of final states which may include many correlated intermediate states. Here, the intrinsic limitations of any
coupled channels approach that is based on ionic bound states only are exposed. Firstly, a correlated intermediate state with a bound character needs large number of ionic states to be
correctly represented. Secondly, the ionic bound states based on gaussian basis sets do not have the exact asymptotic behavior. This can lead to an inaccuracy in
length gauge dipole matrix elements. Finally, the absence of ionic continuum states in our approach can lead to another source of inaccuracy. Due to these limitations, we do not expect the
shake-up channel spectra to be more accurate than 10-15\%.

\myfig
{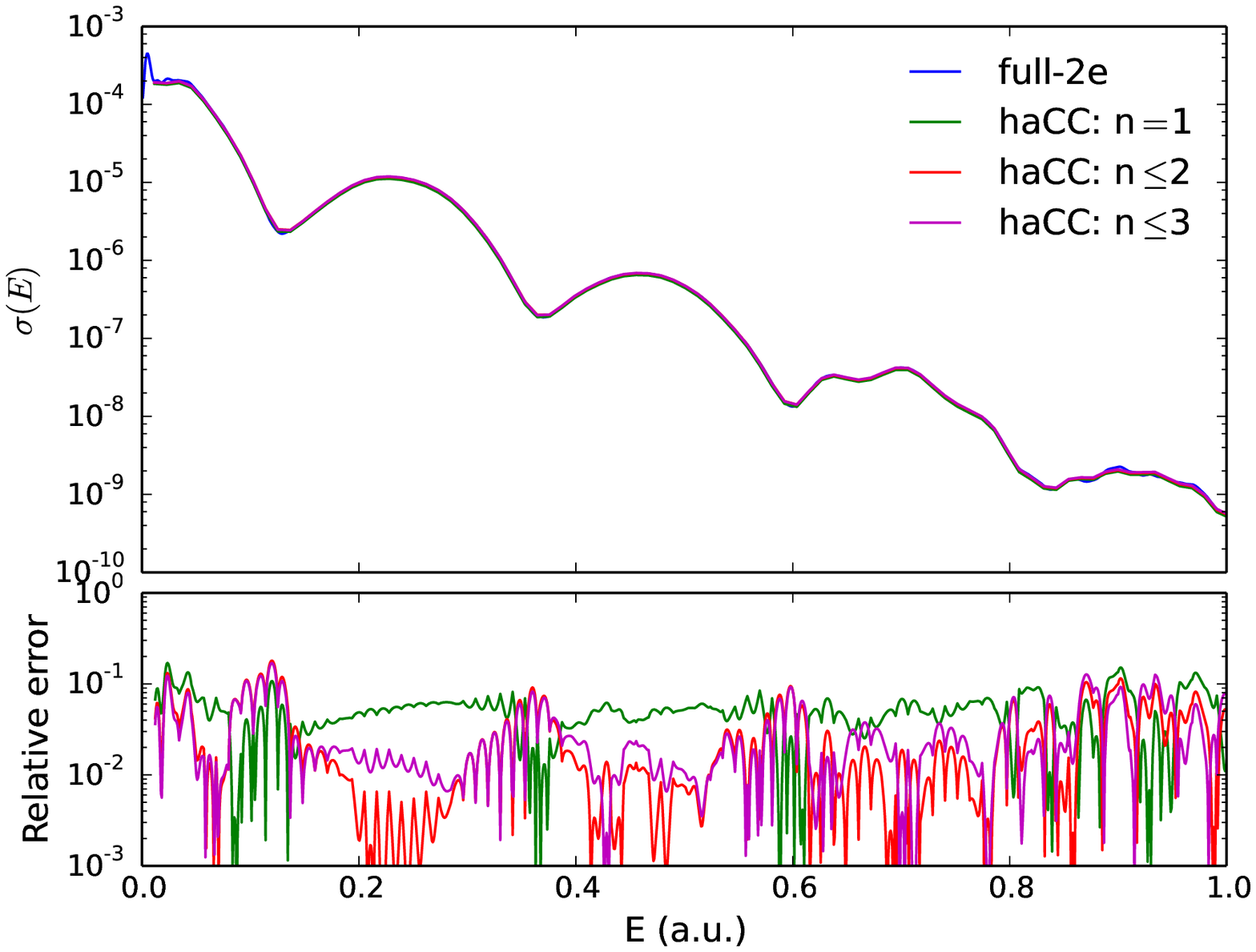} {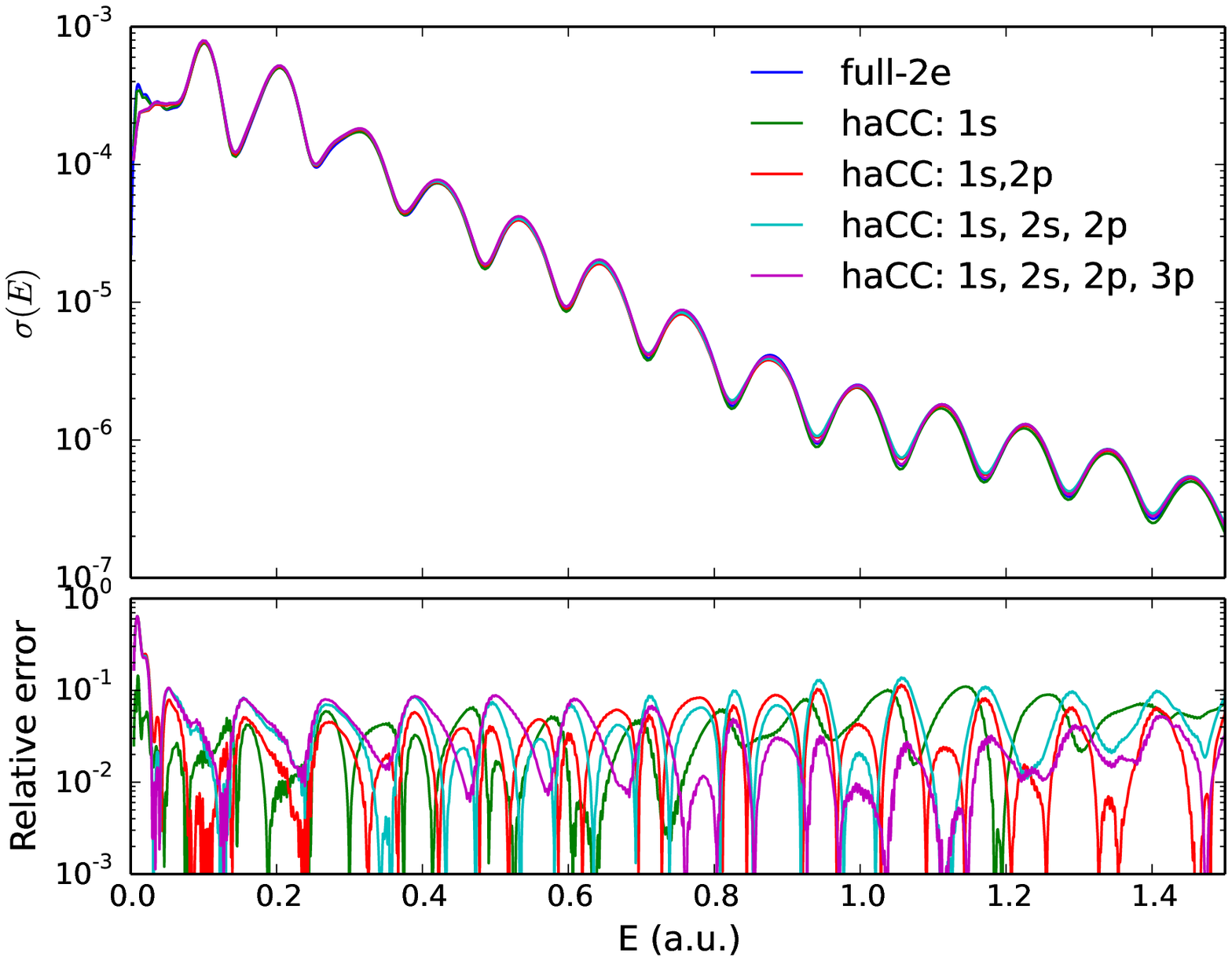} 
{Total photoelectron spectra from helium with Left figure: 3-cycle, 200nm laser pulse with a peak intensity of $10^{14} W/cm^2 $, Right figure: 3-cycle, 400nm
laser pulse with a peak intensity of $3 \times 10^{14} W/cm^2 $. The upper panels  show spectra obtained with a full-2e and haCC calculations with different number
of ionic states included as indicated in the legend. Here, n is the principal quantum number. The lower panels show relative errors of haCC calculations with respect to full-2e
calculation}
{fig:He:200nm,400nm}

Figure \ref{fig:He:200nm,400nm} shows total photoelectron spectra from helium at 200nm and 400nm wavelengths. The exact laser parameters are indicated in the figure captions. At 200nm 
($\omega$ = 0.228a.u), the ionization threshold is four photons. A single ionic state calculation produces spectra that are 10\% accurate with respect to a full dimensional calculation.
Addition of second and third shell ionic states improves the accuracy of the spectra to few percent level in the important regions of the spectrum. At 400nm ($\omega$ = 0.114a.u), the
ionization threshold is 8 photons. Also here, a single ionic state computation produces spectra that are 10\% accurate with respect to a full dimensional calculation. Addition of more 
ionic states, does not improve the accuracy further. This is possibly due to the missing continuum of the second electron that is needed to fully describe the polarization of the ionic core.

At longer wavelengths, we find that single ionic state computations are sufficient to produce spectra accurate on the level of 10\%. This is consistent with the knowledge that at
longer wavelengths, it is the ionization thresholds that play a more important role in determining the ionization yields compared to the exact electronic structure. Our findings
show that helium at long wavelengths can be approximated as a single channel system.

\subsection{Beryllium}
Beryllium is a four electron system that is often treated as a two electron system due to the strong binding of its inner two electrons. The third ionization potential of beryllium is 
153.8961 eV \cite{Drake1988}. With photon energies below this third ionization potential, it can be safely treated as an effective two electron system. This allows us to have a benchmark
for our spectra by adapting the simple Coulomb potential to an effective potential in our two electron code. We use the effective potential given in \cite{PhysRevA.88.053412} for our
benchmark calculations. We refer to these as `effective-2e' calculations.

Table \ref{tab:Be:ene} lists the energies of the first 8 ionic states of beryllium relative to the ground ionic state. Unlike in helium, the ionic states are closely spaced in energy
and one would expect inter-channel couplings to play a greater role.

\begin{table}[ht]
\begin{center}
\begin{tabular}{|c|c|c|}
\hline
Ionic state & NIST database (eV) & Columbus energies (eV)\\
\hline
$1s^2 2s$ & 0.0 & 0.0\\
$1s^2 2p$ & 3.9586 & 3.9767\\
$1s^2 3s$ & 10.9393 & 10.9851\\
$1s^2 3p$ & 11.9638 & 12.1407\\
\hline
\end{tabular}
\caption{Energies of the used single ionic states of Beryllium relative to the ground state ion. The COLUMBUS\cite{Lischka2011} ionic states are computed at MR-CISD level with
aug-cc-pvtz basis.}
\label{tab:Be:ene}
\end{center}
\end{table}

Figure \ref{fig:Be:21nm,200nm} shows photoelectron spectra from beryllium with 21nm and 200nm wavelength laser pulses. The exact parameters are indicated in the figure caption. 
The relative errors of spectra from the haCC calculations are computed with respect to the effective-2e calculations.

At 21nm, the one and two photon ionization peaks of ground state channel spectra are shown. Here, the single photon ionization process itself needs more than the ground ionic state to
produce accurate photoelectron spectra. Adding more ionic states improves the accuracy to a few percent level. We find that the close energetic spacing of beryllium ionic states,
leads to a greater possibility of inter-channel coupling that is manifested in the form of the single ionization continuum needing more than the ground ionic state to be well represented.

\myfig
{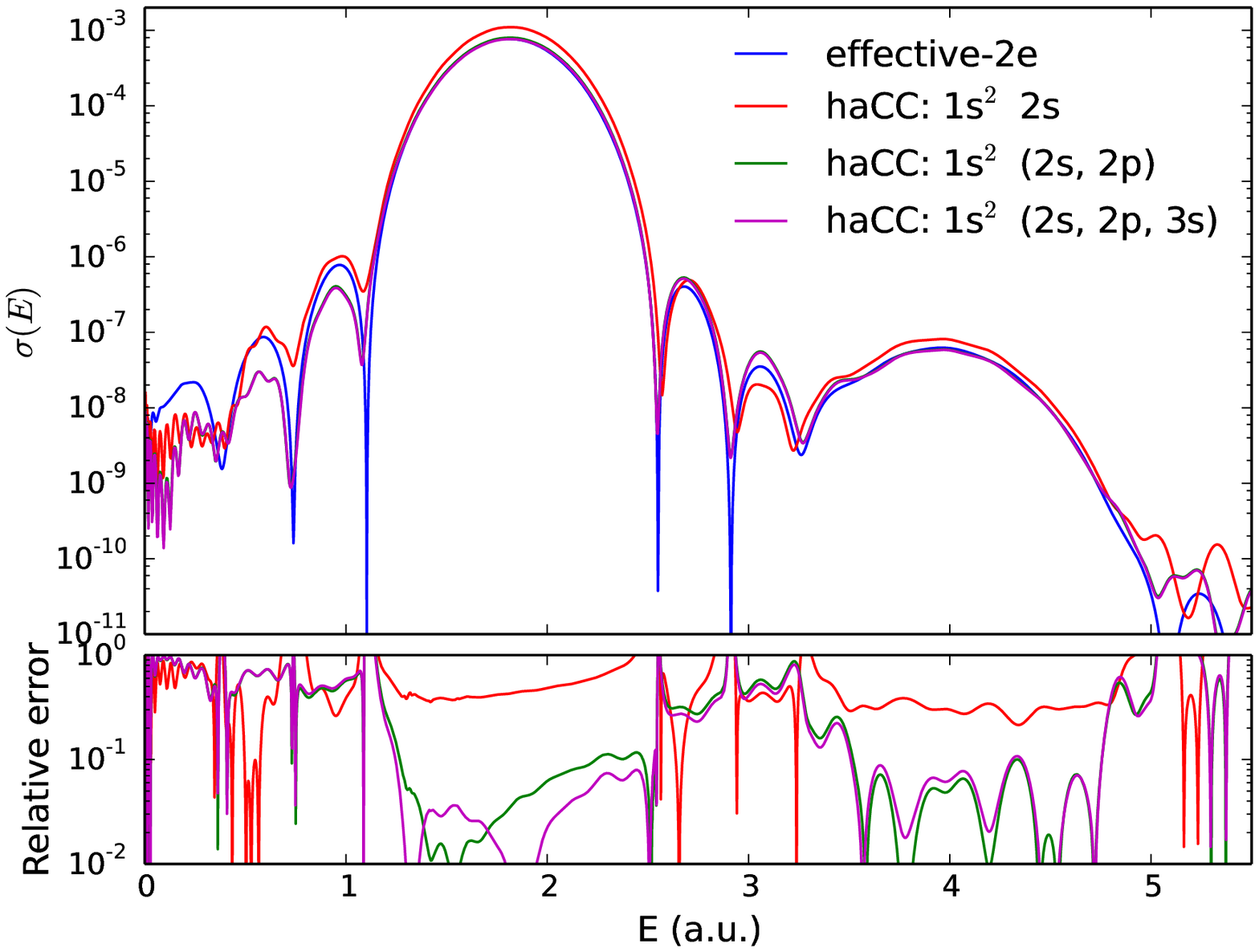} {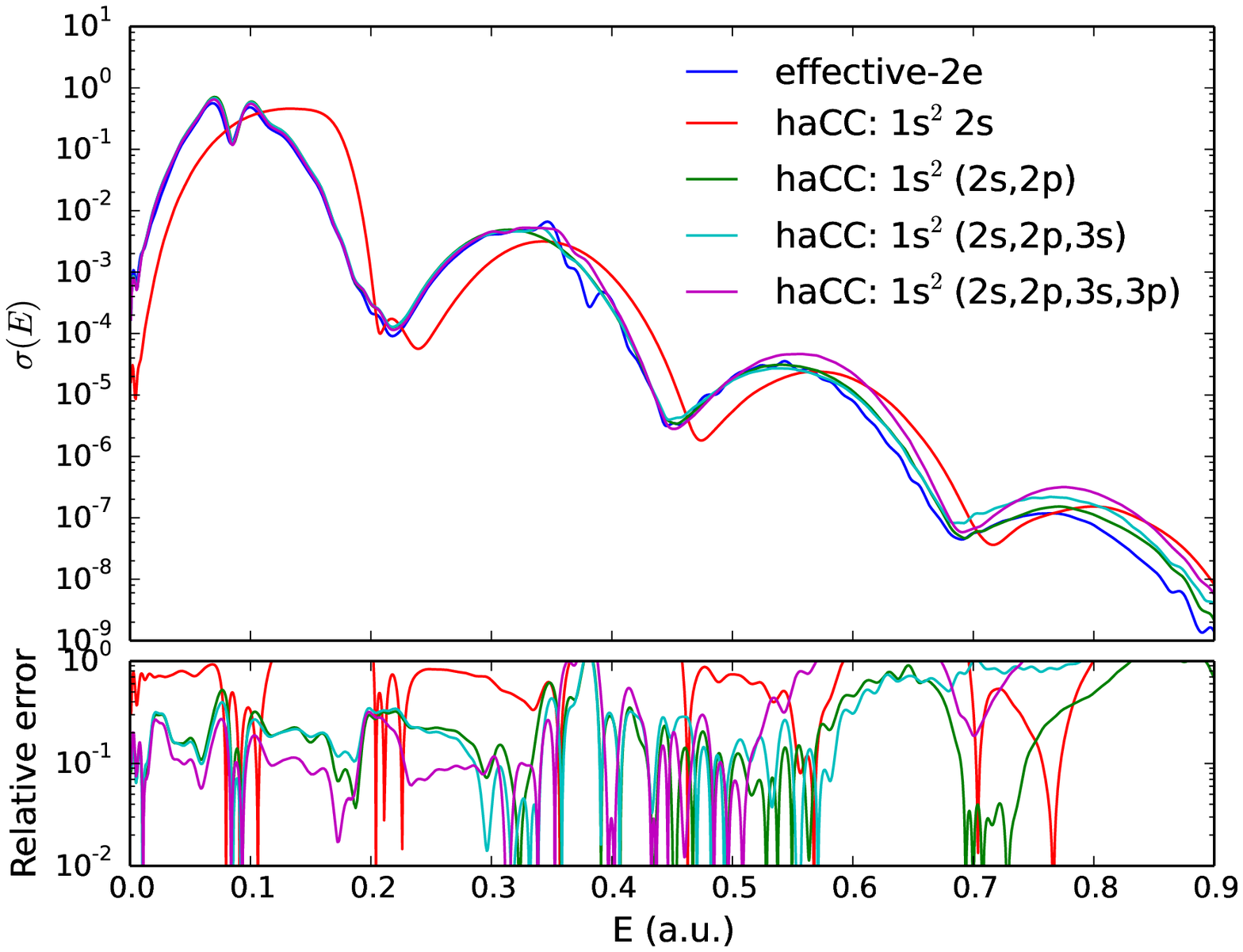}
{Photoelectron spectra from the beryllium atom. Left figure: Ground state channel spectra with 3-cycle, 21nm laser pulse with a peak intensity of $10^{15} W/cm^2 $. Right figure: Total spectra with 3-cycle, 
200nm laser pulse with a peak intensity of $10^{14} W/cm^2 $. The upper panels show spectra obtained with effective-2e and haCC calculations with different number of ionic states included
as indicated in the legend. The lower panels show relative errors of haCC calculations with respect to the effective-2e calculations.}
{fig:Be:21nm,200nm}

Also at 200nm, we need more than the ground ionic state to compute realistic spectra. With $1s^2 2s$, $1s^2 2p$ ionic states included, the spectra produced have 20\% accuracy
with respect to the benchmark calculation. With the addition of $1s^2 3s$ and $1s^2 3p$ states, a structure similar to the one predicted by the benchmark calculation develops around 10 eV. 
This structure may be identified with the lowest resonance $1s^2 2p 3s$ at 10.71 eV \cite{PhysRevA.88.053412}. The coupled channels calculations with the number of ionic states considered here, 
however do not reproduce the structure on the second peak exactly. This points to a feature of a coupled channels basis that, the correct representation of a strongly correlated state that
has bound character requires a large number of ionic states. As an alternative strategy, one can explicitly include the correlated state of importance into the basis, if it can be
pre-computed, on the same footing as the correlated ground state.

It has been shown through examples in section \ref{sec:helium} that helium can be modeled as single channel system at longer wavelengths. Lithium, the smallest alkali metal, also has been
successfully modeled as a single electron system in an effective potential, for example in \cite{PhysRevA.87.063405}. We find that beryllium needs at least two ionic states - $1s^2 2s$
and $1s^2 2p$ for a realistic modeling. It serves as a first simple example where single electron models break down and multiple channels need to be considered.

\begin{figure}[ht]
\centering
 \includegraphics[scale=0.5]{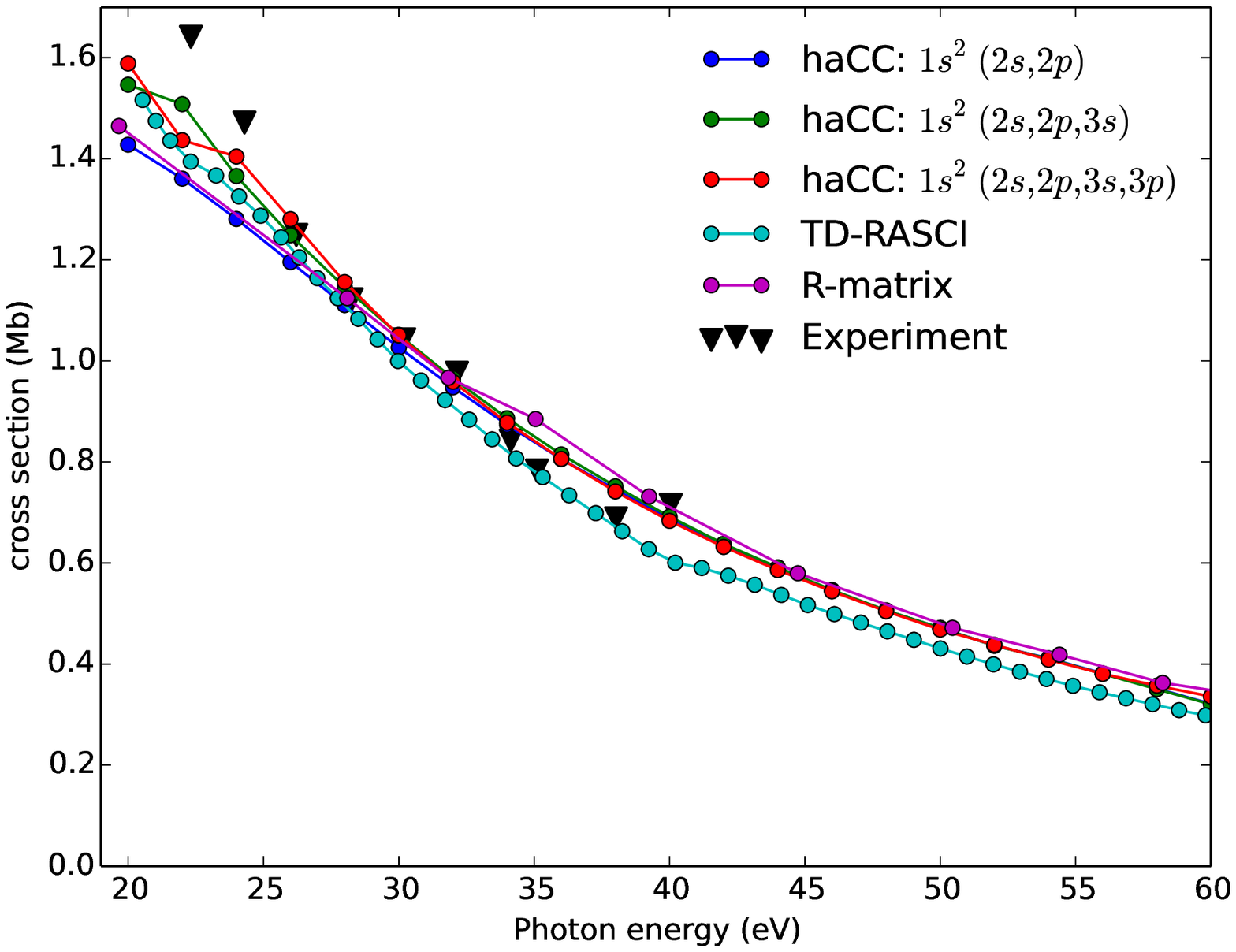}
 \caption{Single photoionization cross-sections for beryllium in the photon range of 20-60 eV. Presented are results from haCC calculations with 4,5,8 ionic states. The figure shows
 a comparison with earlier calculations using TD-RASCI method \cite{PhysRevA.86.053424}, R-matrix method \cite{Berrington1997} and with experimental results from \cite{PhysRevA.71.012707}}
 \label{fig:Be:1phscan}
\end{figure}

In figure \ref{fig:Be:1phscan}, we present the single photon ionization cross-sections as a function of photon energy from our haCC method and compare them with the cross-sections
calculated with TD-RASCI method \cite{PhysRevA.86.053424} and R-matrix method \cite{Berrington1997} and with experimental results from \cite{PhysRevA.71.012707}. The cross-sections in our time
dependent approach are computed using the Eq (51) given in ref \cite{PhysRevA.79.053411} with which the N photon ionization cross-section, $\sigma^{(N)}$ in units $cm^{2N}/s^{N-1}$ can be computed as:

\begin{equation}
 \sigma^{(N)} = \left( 8 \pi \alpha \right)^N \left( \frac{3.5 \times 10^{16}}{I}\right)^N \omega^N \Gamma a_0^{2N} t_0^{N-1}
\end{equation}
where $I$ is the intensity in $W/cm^2$, $\omega$ is the laser frequency in a.u, $\alpha$ is the fine structure constant and $a_0$, $t_0$ are atomic units of length and time respectively.
$\Gamma$ is the total ionization rate which is computed in a time dependent approach by monitoring the rate at which the norm of the wavefunction in a certain inner region drops. We use
for our computations a 40-cycle continuous wave laser pulse with a 3 cycle $\cos^2$ ramp up and ramp down and with an intensity of $10^{12} W/cm^2$.

All the theoretical methods agree with each other qualitatively, though there are differences on the level of 5-10\% quantitatively. The experimental results from \cite{PhysRevA.71.012707} 
have error bars on the level $\lesssim$ 10\% (0.1 Mb) which are not shown here. All the theoretical results lie in this range except at low energies. In the higher photon energy range, 30-60 eV, the 
haCC results and the R-matrix results are in good agreement compared to the TD-RASCI. In the haCC calculations, including more than 4 ionic states does not change the cross-sections.
In the photon energy range 20-30 eV, the haCC computations with 5 and 8 ionic states are in good agreement with TD-RASCI results compared to the R-matrix results. In this energy range, 
the cross-sections from haCC calculations show a dependence on the number of ionic states included. This modulation may be attributed to the presence of auto-ionizing states in 
this region. Table III in \cite{PhysRevA.88.053412} presents a list of resonances that appear in beryllium electronic structure. The first ionization potential is 9.3 eV. With photon energies
around 20 eV, the resulting photoelectron reaches continuum region where a number of resonances are present. As correlated resonances need many ionic states to be well represented in a
coupled channels basis, this may explain the dependence of the cross-section on the number of ionic states in 20-30 eV photon range.

\subsection{Hydrogen molecule}
The hydrogen molecule in linearly polarized laser fields parallel to the molecular axis, with fixed nuclei has the same symmetry as helium in linearly polarized laser fields. The off-centered
nuclear potential, however, increases the angular momenta requirement when treated with a single center expansion. While the number of required basis functions can be reduced through
a choice of a more natural coordinate system like prolate spheroidal coordinates for diatomics \cite{Vanne2004}, the challenge of computing two electron integrals remains. In the case of
hydrogen molecule at equilibrium internuclear distance ($R_0$ = 1.4 a.u), a calculation with single center expansion easily converges, as the light hydrogen nuclei do not significantly
distort the spherical symmetry of the electron cloud. As a benchmark for spectra, we use results from a full dimensional calculation, that expands the wavefunction in a single center basis.

\myfig
{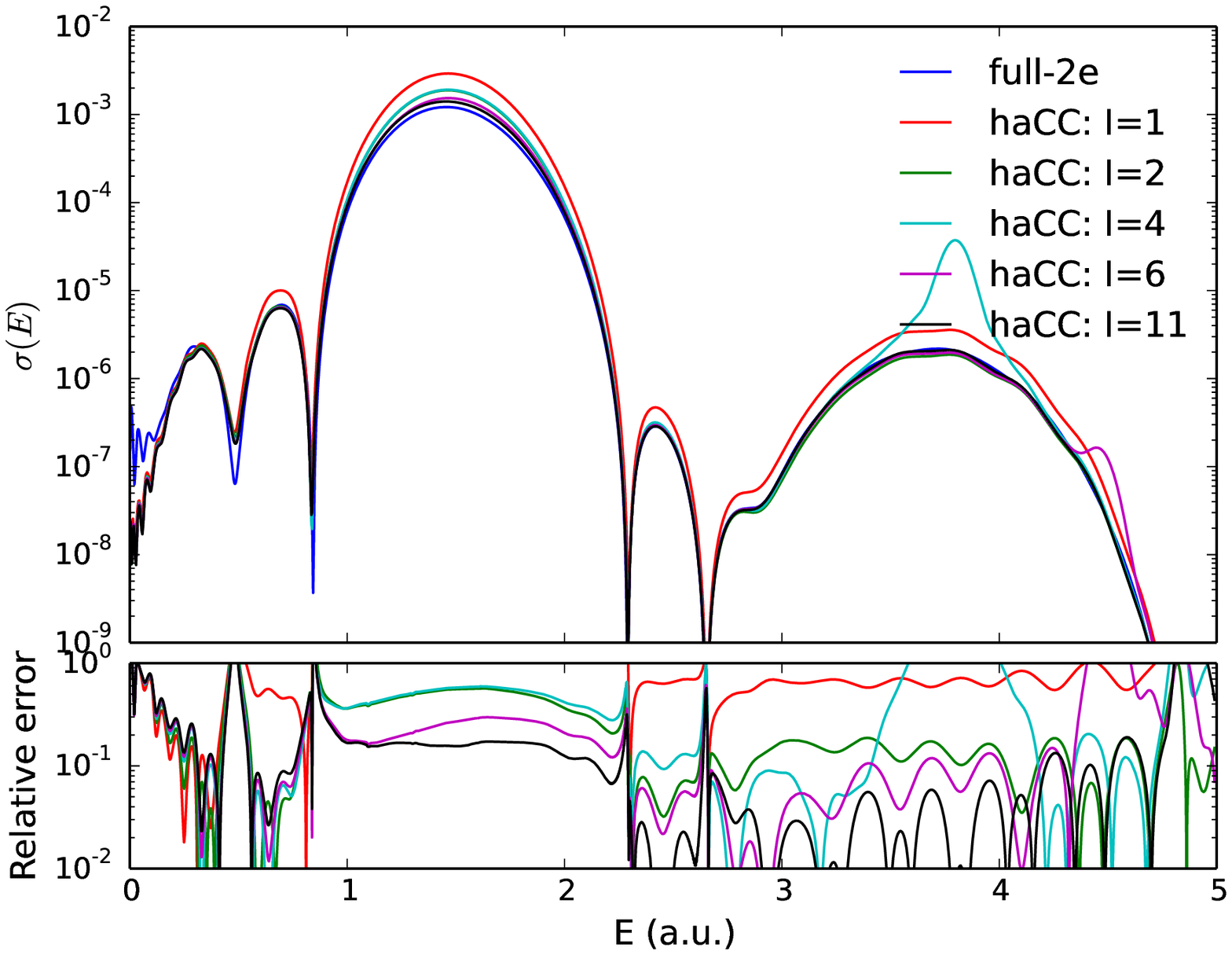} {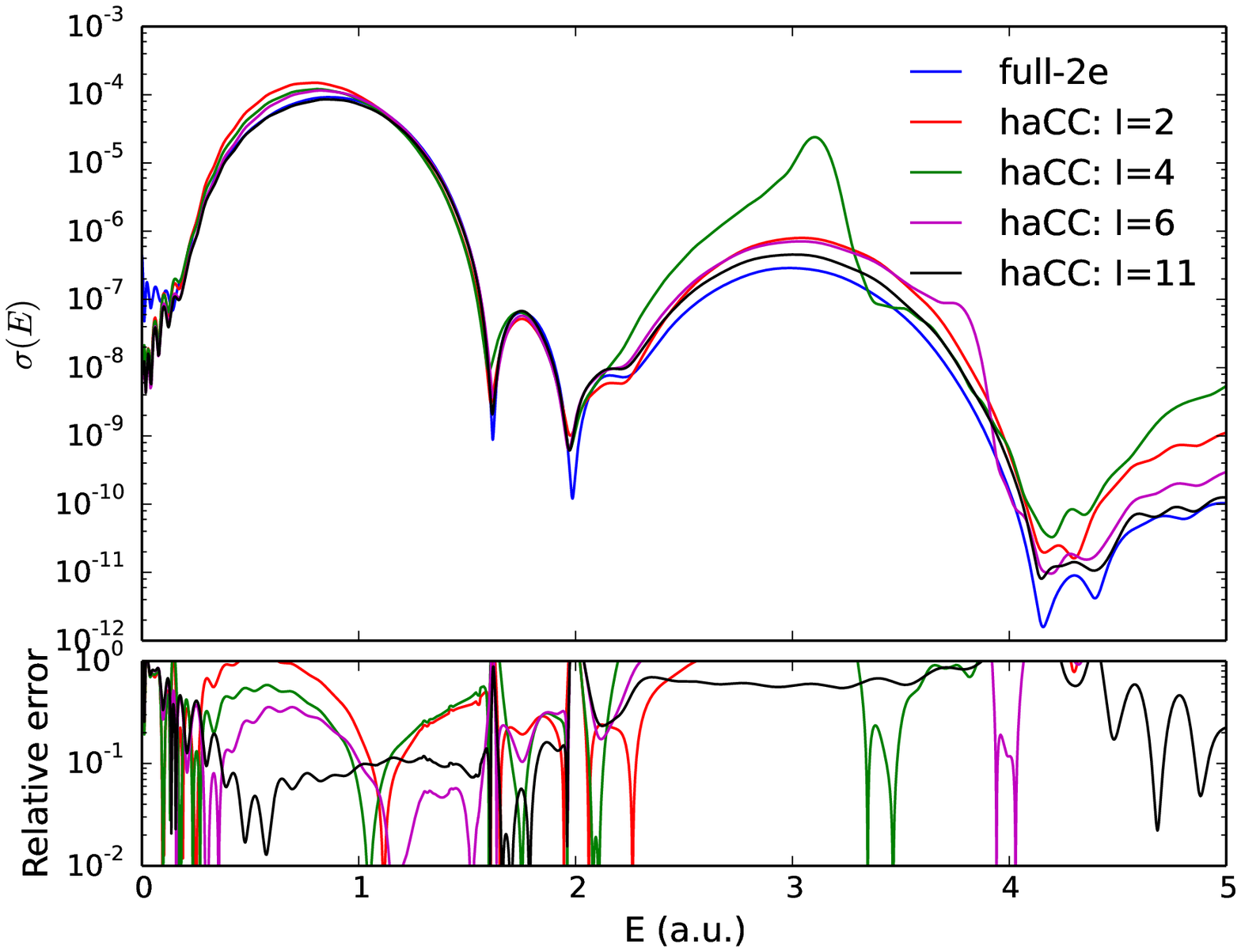}
{Photoelectron spectra from $H_2$ with a 3-cycle 21nm laser pulse with a peak intensity of $10^{15} W/cm^2 $. Left figure: Ground state channel ($1\sigma_g$) Right figure: First excited state channel
($1\sigma_u$). The upper panels show spectra obtained with full-2e and haCC calculations with different number of ionic states (I) included (as indicated in the legend). 
The lower panels show relative errors of haCC calculations with respect to the full-2e calculation. With I=4,6, there are visible artefacts on the 2 photon peaks around 3 a.u. which are explained
in the text.}
{fig:H2:21nm}

Figure \ref{fig:H2:21nm} shows photoelectron spectra from $H_2$ at 21nm wavelength. The exact laser parameters are given in the figure caption. The ground state ($1\sigma_g$) and 
first excited state ($1\sigma_u$) channel spectra are shown. We find that, at this wavelength, a single ionic state is not sufficient to produce accurate photoelectron spectra. With the addition of more ionic states, there
is a systematic improvement in the accuracy of the calculations. With 11 lowest $\sigma$ and $\pi$ ionic states included, we obtain an accuracy of about 10\% for the $1\sigma_g$ channel. The 
single photon ionization to the shakeup channel $1\sigma_u$ is also computed to a few percent accuracy with 11 ionic states. We find that the single ionization continuum of $H_2$ is more
complex unlike helium and it needs more than a single ionic state.

With 4 and 6 ionic states, we find artefacts on the two photon peaks. This is a result of a part of the quantum chemistry neutral that appears in the eigenvalue spectrum of the 
Hamiltonian as a spurious doubly excited state ($|s\rangle$). Let $|f_i\rangle$ be the eigenvectors of the N particle field free Hamiltonian to which $|\mathcal{G}\rangle$ contributes. We 
can find a state $|s\rangle$ such that
\begin{equation}
 |s\rangle = | \mathcal{G} \rangle - \sum_i | f_i \rangle \langle f_i | \mathcal{G} \rangle \neq 0
\end{equation}
 This spurious correlated state moves to higher with addition of ionic states. A straight forward solution to this problem is to compute this state and project
 it out of the basis. But, this would require locating the spurious state in the eigenvalue spectrum, which is very demanding for large Hamiltonians. Fortunately, by their dependence on the number
 of ionic states, artefacts of this kind are easily detected and can be moved out of the region of interest by using sufficiently many ionic states. This is a possible artefact one needs
 to be cautious about when dealing with wavefunction ansatz of kind (\ref{eq:ansatz}).

 \myfig
{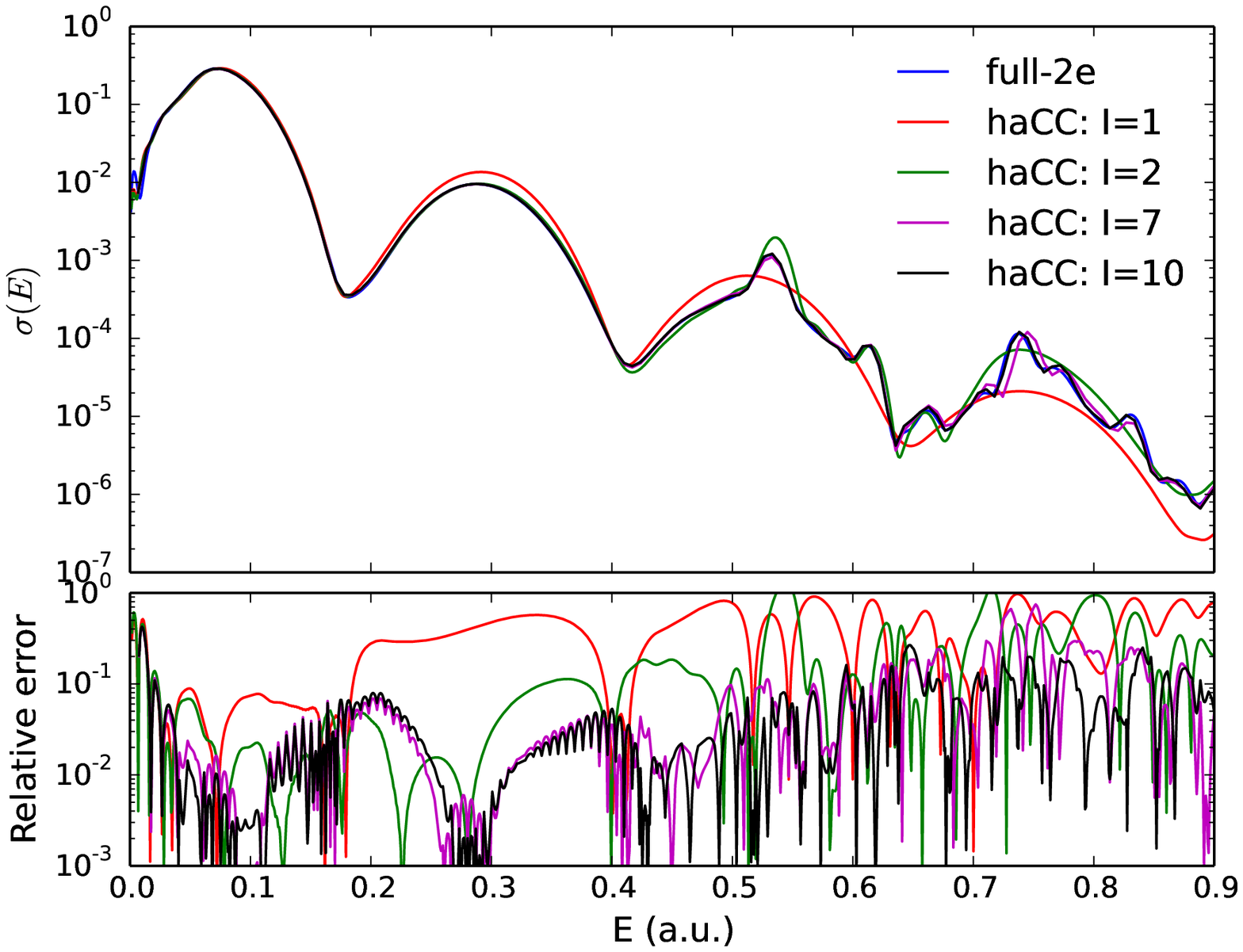} {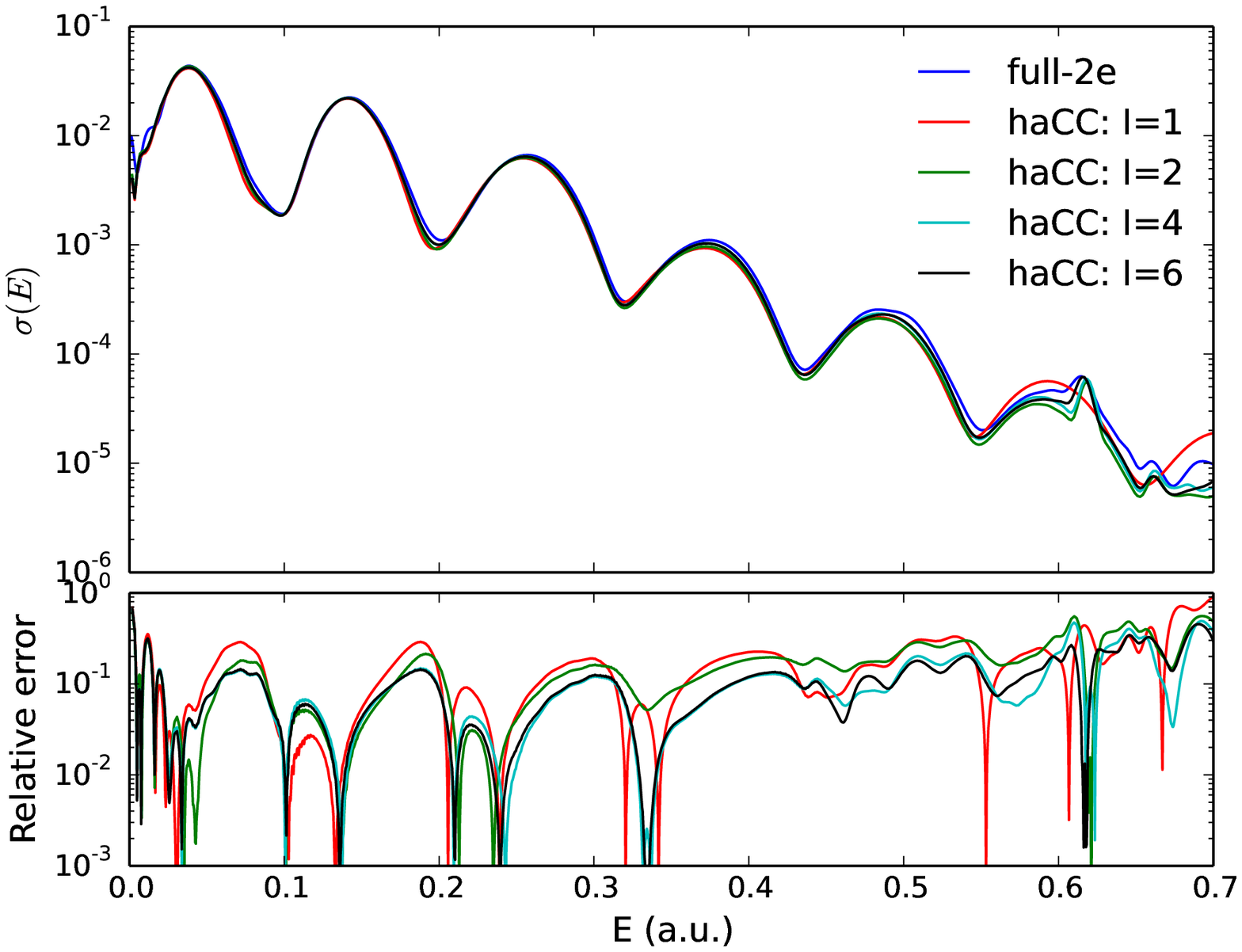}
{Total photoelectron spectra from $H_2$ with - Left figure: 3-cycle 200nm laser pulse with a peak intensity of $10^{14} W/cm^2 $. Right figure: 3-cycle 400nm laser pulse with a peak intensity
of $10^{14} W/cm^2 $. The upper panels show spectra obtained with full-2e and haCC calculations
with different number of ionic states (I) included as indicated in the legend. The lower panels show relative errors of haCC calculations with respect to the full-2e calculation.}
{fig:H2:200nm,400nm}
 
Figure \ref{fig:H2:200nm,400nm} shows total photoelectron spectra at 200nm and 400nm wavelengths. At 200nm, spectra are accurate up to 10\% with 2 ionic states. Addition of more ionic states
helps reproduce additional resonant features in the spectrum. Also at 400nm, 2 ionic states are sufficient to compute spectra that are accurate on 10\% level, except for the resonant features. 
Inclusion of up to 6 ionic states reproduces the feature around 0.62 a.u. This feature in the 400nm spectrum may be identified with second or third $^1\Sigma_u^+$ doubly excited state 
\cite{sanchez1997}.

We find that with $H_2$ at longer wavelengths, ground ionic state is sufficient to compute realistic spectra excluding the resonant features. Addition of more ionic states produces
the additional features in the spectrum resulting from presence of correlated doubly excited states.

\begin{figure}[ht]
 \centering
 \includegraphics[scale=0.5]{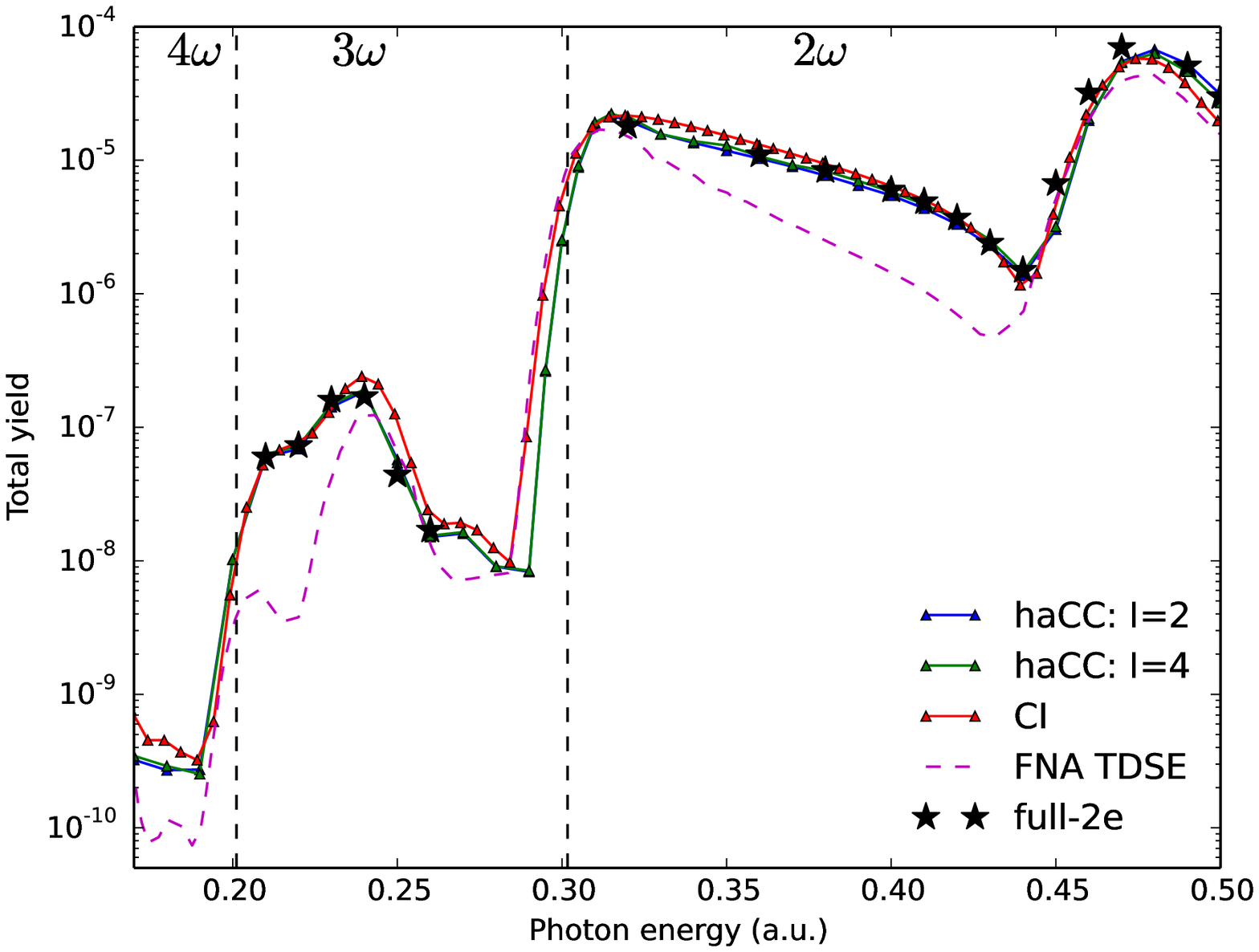}
 \caption{Ionization yield from $H_2$ at equilibrium internuclear distance ($R_0 = 1.4 a.u$) as a function of photon energy. Laser parameters: $10^{12} W/cm^2$ peak intensity,
 $\cos^2$ envelope pulses and 10fs pulse duration (In Eq (\ref{pulse}) $2cT = $ 10fs).
 A comparison of haCC calculations with 2,4 ionic states with CI results from \cite{PhysRevA.90.053424} and FNA-TDSE results from \cite{PhysRevLett.96.143001}. The dashed vertical lines 
 separate different multi-photon ionization regimes.}
 \label{fig:H2:wvscan}
\end{figure}

Figure \ref{fig:H2:wvscan} shows total ionization yield as a function of photon energy in the range 0.17-0.5 a.u. It shows a comparison of our haCC calculations with data available from 
other theoretical methods - time dependent CI method from \cite{PhysRevA.90.053424} and FNA-TDSE (fixed nuclei approximation) method from \cite{PhysRevLett.96.143001}. The vertical lines in the 
figure separate different multi-photon ionization regimes. The general behavior of the yields from both CI and haCC procedure is the same. There is a discrepancy in the yields on the
level of 15\%. This may be a result of the intrinsic limitations of both the methods. There is a shift of about 0.2eV between the CI and haCC results around 0.3 a.u photon energy.
The ionization potential of $H_2$ at equilibrium inter-nuclear distance ($R_0 = 1.4 a.u$) is 0.604492 a.u (Table I in \cite{PhysRevA.78.042510}). The ionization potential in our calculations
is 0.6034 a.u, while it is around 0.59 a.u in time dependent CI calculations from \cite{PhysRevA.90.053424} which can be read from the multi-photon thresholds presented there. Judging from
the accuracy of the ionization thresholds, one may conclude that our haCC computations are more accurate than the CI results. The FNA-TDSE results from \cite{PhysRevLett.96.143001} have a 
similar qualitative behavior but are quantitatively quite different from our results and the CI results. The full-2e computations are expensive, especially at the ionization thresholds where large
simulation box sizes are needed due to excitation of Rydberg states. Hence, we used this method to compute only at few points specially in the range where the FNA-TDSE and the haCC results
have large discrepancies. We find that the full-2e calculations are consistent with the haCC results supporting the accuracy of our computations.

\section{Conclusions}

The hybrid anti-symmetrized Coupled Channels method introduced here
opens the route to the reliable {\it ab initio} calculation of fully 
differential single photo-emission spectra from atoms and small molecules 
for a broad range of photon energies. It unites advanced techniques for the solution of the 
time-dependent Schr\"odinger equation for one- and two-electron systems 
in strong fields with state of the art quantum chemistry methods for the 
accurate description of electronic structure and field-induced bound state 
dynamics. For the specific implementation we have relied on a finite element
implementation of the strong field dynamics and Gaussian-based CI package of 
COLUMBUS. 

Key ingredients for the successful implementation are good performance of 
tSURFF for the computation of spectra from comparatively small spatial domains
on the one hand and access to the well established technology 
of quantum chemistry on the other hand. We could obtain the quantum 
chemical structure in the form of the complete expansion into determinants from COLUMBUS. 
In future implementations, it may be sufficient to output from a given package
the generalized one, two, and three-electron density matrices together with 
generalized Dyson orbitals, both defined in the present paper. It turned out to 
be instrumental for accurate results that haCC allows for the inclusion 
of neutral states in a natural fashion and at very low computational cost.

Several new techniques were introduced and implemented for the establishment of the 
method. Most notably, the mixed gauge approach \cite{mixedg} turned out to be crucial 
for being able to take advantage of the field-free electronic structure in presence of
a strong field without abandoning the superior numerical properties of a velocity-gauge 
like calculation. The finite element method used for single-electron strong-field dynamics
is convenient, but certainly not the only possible choice. Similar results should be
achievable with higher order B-spline methods or any other discretization suitable for
solutions of the single electron strong field Schr\"odinger equation. Low-rank updates
are used in several places for the efficient computation of the inverses of the large overlap
matrix and to control the linear dependency problems arising from anti-symmetrizing the essentially 
complete finite elements basis against the Hartree-Fock orbitals. 

We have made an effort to explore the potential range of applicability of the method
by performing computations in a wide range of parameters on a few representative
systems, where results can be checked against essentially complete methods.
Spectra for the He atom were independently obtained from fully correlated 
two-electron calculations. We could demonstrate that haCC gives spectra on 
the accuracy level 10\% with very low effort. An interesting observation is that
in the long wavelength regime indeed a single ionization channel produces
correct results, justifying {\it ex post} wide spread model approaches of the 
strong field community. As a note of caution, we recall that this is only 
possible as the {\em fully correlated initial state} is routinely included in 
the haCC scheme. At short wavelength, the ionic excited state dynamics plays 
a larger role and reliable results require inclusion of up to 9 ionic channels.
With this we could correctly resolve also the peak due to He's doubly excited state.

The second atomic system, Be, was chosen to expose the role of ionic dynamics. While the 
$1s$ core electrons are energetically well-separated and no effect of their dynamics
was discernable in a comparison with a frozen core model, the narrow spaced 
ionic states preclude single channel models. Depending on the observable and 
on desired accuracies, a minimum of two ionic channels had to be used.

For the comparison of $H_2$ photoionization and photoelectron spectra, we could 
refer to literature and supplemented the data with a full two-electron calculation.
At 400nm, $H_2$ can be treated as a single channel system. At intermediate wavelengths, we find
the need for at least two ionic channels, and we could obtain a fair agreement 
with comparison data. Here one has to take into consideration that all alternative methods
operate near the limits of their applicability.

With this set of results we demonstrated the correctness of the method and its
essential features. In our calculations, also the fundamental limitations of the approach
were exposed. Clearly, the field-induced dynamics of the ionic part must be describable 
by a few states with bound character. haCC shares this limitation with any expansion
that is limited to a few ionic states. Note that the problem is partly mitigated by 
the possibility to include fully correlated ground as well as singly- and doubly-excited 
states with bound-state character that are known to appear in the dynamics.
  
The method in its present implementation can be applied to small molecules such 
as $N_2$ and $CO_2$ which will be reported in a forthcoming publication. At the moment,
the computation of the two-electron integrals poses a mild technical limitation for
such calculations, and an improvement of the presently rather straight-forward algorithm
is needed for going to larger systems. Another limitation arises when the molecule becomes
too large for computing even strong field single-electron dynamics over its complete extension. 
At present, tSURFF allows us to limit computation boxes to the scale of $\sim 40\, a.u.$.
Also, for the single electron part, we use at present single-center expansions, which perform
notoriously poorly if scattering centers are distributed over more than a few atomic units.
This limitation may well be overcome by a more versatile single-electron discretization,
though at significant implementation effort.

Other potential extensions are to double-emission. The tSURFF method was formulated for this situation.
Combining such already sizable calculations with a dication described by quantum chemistry in the same
spirit as here may be feasible. The formula presented can be readily extended to include that case. However,
the scaling is poor such that one may only hope for the simple one- or two-channel 
situation to be feasible in practice. A cut-down version of such an approach can be used to 
include non-bound dynamics by describing a second electron's dynamics in a more flexible basis,
however, without admitting its emission.

These lines of development will be pursued in forthcoming work.

\section*{Acknowledgements}
The authors thank the COLUMBUS developers - Hans Lischka, University of Vienna; Thomas M\"uller, Forschungszentrum J\"ulich; Felix Plasser, University of Heidelberg and Jiri Pittner,
J. Heyrovsk\'y Institute for their support with constructing the quantum chemistry interface. V.P.M. is a fellow of the EU Marie Curie ITN `CORINF' and the International Max Planck
Research School - Advanced Photon Science. A.Z acknowledges support from the DFG through excellence cluster `Munich Center for Advanced Photonics (MAP)' and from the Austrian Science
Foundation project ViCoM (F41).
 
\appendix

\section{Finite element basis} \label{sec:apdx_fem}
Let $\{r_0,r_1,...,r_{n}\}$ be points on the radial axis that define $n$ intervals on the radial axis. In a finite element approach, the basis functions $f_i^n(r)$ are chosen such that 
\begin{equation}
 f^{n}_{i}(r) 
 \begin{cases}
    \neq 0  & \text{if } r \in [r_{n-1},r_n]\\
    = 0 & \text{otherwise}
\end{cases}
\end{equation}
The individual basis functions can be chosen from any complete space, for example, in our case we use scaled Legendre polynomials of typical orders 10-14. Here, we write the finite element
index and the function index separately to emphasize that we have two convergence parameters: the order and the number of finite elements. The calculations converge
quickly with increasing order compared to with increasing number of elements \cite{elander1993}. The basis functions should also be tailored to satisfy the continuity conditions.
This may be accomplished through an orthogonalization procedure in each interval so that the functions on each interval satisfy the following conditions:
\begin{equation}
\begin{aligned}
 f^{n}_{0}(r_{n-1}) = 1; \;\; & f^{n}_{0}(r_{n}) = 0 \\
 f^{n}_{1}(r_{n-1}) = 0; \;\; &  f^{n}_{1}(r_{n}) = 1 \\
 f^{n}_{i\neq0,1}(r_{n-1}) = 0; \;\; &  f^{n}_{i\neq0,1}(r_{n}) = 0 
 \end{aligned}
\end{equation}
Even though we are solving a second order differential equation, it is sufficient to impose just the continuity condition to solve the differential equation. It can be shown through
a simple computation, for example as shown in \cite{elander1993}, that the matrix elements corresponding to the laplacian operator can be computed even if the functions are not differentiable
at the finite element boundaries. This is because the $\delta$-like terms arising due to the non-differentiability are compensated by the surface terms. The matrices corresponding to various
operators in a finite element basis have a banded structure, that can be used to perform various linear algebra operations efficiently.

In a three-dimensional situation with spherical symmetry, these radial finite element functions can be multiplied by a complete set of angular basis functions such as the spherical harmonics
to construct a three dimensional basis of the form $f_i^n(r) Y_{lm}(\theta,\phi)$.

\section*{References}
\bibliography{paper_refs}{}
\bibliographystyle{unsrt}

\end{document}